\documentstyle [12pt,epsfig]{article}
\begin{document}
\def\slash#1{\not\! #1}
\newcommand{\bfk}{\mbox{\bf k}}
\thispagestyle{empty}
\begin{flushright}
LPC 96 53 \\
INFNCA-TH9620 \\
hep-ph/9611337 \\
November 1996
\end{flushright}
\vspace{2.5truecm}
\begin{center}
{\Large \bf{Production of meson pairs, involving tensor \\
and pseudotensor mesons, in photon-photon collisions}} \\[10pt]
\vspace{18pt}
Ladonne Houra-Yaou, Paul Kessler, and Joseph Parisi \\
{\it Laboratoire de Physique Corpusculaire, Coll\`ege de France \\
11, Place Marcelin Berthelot, F-75231 Paris Cedex 05, France}\\[5pt]
\vspace{8pt}
Francesco Murgia \\
{\it Istituto Nazionale di Fisica Nucleare, Sezione di Cagliari \\
Via Ada Negri 18, I-09127 Cagliari, Italy} \\  
\vspace{8pt}
Johan Hansson \\
{\it Department of Physics, Lule\aa{} University of Technology,
S-95187 Lule\aa, Sweden}
\end{center}

\vspace{2truecm}

\begin{center}
{\bf ABSTRACT}
\end{center}
\vspace{16pt}

Starting from a bound-state model of weakly bound quarks for
($q\,\bar{q}$) mesons, we derive a formalism for computing
the production or decay of such mesons, whatever the value
of their internal orbital angular momentum $L$. That
approach appears as a natural generalization of the
Brodsky-Lepage formalism (valid only for $L=0$)
that has been widely used in recent
years for the computation of exclusive processes in
perturbative QCD. We here apply it to the production,
in photon-photon collisions, of~:
{\it i}\,) tensor-meson pairs~; {\it ii}\,) pseudotensor-meson
pairs~; {\it iii}\,) hybrid pairs made of a pion and a
pseudotensor meson. The numerical results we obtain allow
for some hope of experimentally identifying such pairs,
in the charged channels, at high-energy $e^+e^-$
colliders of the next generation, provided integrated
luminosities as high as $\approx 10^{40}$ cm$^{-2}$
can be reached.

\clearpage

\section{Introduction}

Some fifteen years ago Brodsky and Lepage presented a formalism,
based on perturbative quantum chromodynamics, that allows for
the calculation of exclusive processes involving particle
production or decay \cite{brle}. Over the years it
has given rise to a number of more or less successful
applications, such as for instance hadron pair production in
photon-photon collisions \cite{brod}. 
However it is well known that, as far as mesons are concerned,
the Brodsky-Lepage formalism can only be applied to ($q\,\bar{q}$)
states with zero orbital angular momentum, i.e. to pseudoscalar
and vector ($q\,\bar{q}$) mesons. In particular, the production or
decay of scalar, axial-vector or tensor ($q\,\bar{q}$) mesons,
which all have their orbital angular momentum $L$ equal to one,
as well as of pseudotensor ($q\,\bar{q}$) mesons with $L = 2$,
cannot be treated in that formalism.

The purpose of this paper is to define a procedure for the
calculation of exclusive processes involving the production or
decay of ($q\,\bar{q}$) mesons with any orbital angular momentum $L$.
Starting from a bound-state model of weakly bound quarks for such
mesons \cite{cahn}, we shall derive an approach that will basically
appear as a natural generalization of the Brodsky-Lepage formalism. 
Actually the formalism here shown is closely related to the one presented
by some of the authors in former papers bearing on the calculation
of processes that involve the production or decay of $(g\,g)$
glueballs \cite{kada,yaou,icho,murg}.
Let us also recall the mathematical framework that was developed
in a similar context by Benayoun and Froissart \cite{bena}.

Section 2 of this paper will be devoted to the derivation
of our formalism. In section 3 that formalism will be applied
to the production, in photon-photon collisions, of~:
(i) $(q\,\bar q)$ tensor-meson pairs~; (ii) $(q\,\bar q)$
pseudotensor-meson pairs~; (iii) hybrid pairs made of a pion
and a $(q\,\bar q)$ pseudotensor meson~; numerical results will be
presented for those three applications. Section 4 contains a
brief conclusion. Details of calculation are given in three 
Appendices.
  
\goodbreak

\section{A formalism for the production of
 ($\mbox{\boldmath $q \bar{q}$}$) mesons with 
 any orbital angular momentum $\mbox{\boldmath $L$}$}

\nobreak

Let us consider a production process $ a b \to Q c $, where $Q$
is a ($q\,\bar{q}$) meson and $a,~b,~c$ are any particles.
This process is represented schematically in Fig.~1,
both in its center-of-mass frame (A) and in the meson rest frame (B).
In the latter the quark $q$ resp. the antiquark $\bar{q}$ is 
moving inside the meson with a Fermi momentum $\bfk$ resp. $-\bfk$. 
Let us notice from the start that the demonstration presented
hereafter can be extended in a trivial way to the inverse
process $Q c \to a b$.

Calling ${\cal M}$ the amplitude of the process $ a b \to Q c $
and ${\cal T}$ that of the corresponding process at the parton
level, i.e. $ a b \to q \bar{q} c $, a prescription given in
Ref.~\cite{cahn} relates ${\cal M}$ and ${\cal T}$ in the
following way
\footnote{\,
  With respect to Eq.~(2.1) of Ref.~\cite{cahn} we here introduce
  a slight modification (replacing the quark and antiquark masses
  by the corresponding energies in the meson rest frame) which has
  no consequence from the point of view of practical applications
  of the formalism here presented.}~:

\begin{eqnarray}
 {\cal M} & = &  \int d^{3}\bfk~
 \Psi^{\textstyle *} (\bfk)~ 
 {\cal T}\, 
 \frac{(2 \pi)^{3/2} (2 M)^{1/2}}
 {(2 \pi)^{3/2}\sqrt{2(m^2+\bfk^2)^{1/2}}\,
 (2 \pi)^{3/2}\sqrt{2(\bar{m}^2+\bfk^2)^{1/2}}}\; ,
 \label{mabqc}
\end{eqnarray}

\noindent
where $ \Psi(\bfk) $ is the Fermi-momentum distribution of the
quark within the meson (or, in other words, the meson wave
function in momentum space, defined in the meson rest frame). 
The last factor at right hand arises from the normalization of
the S matrix. $M,~m$ and $\bar{m}$ are, respectively, the masses 
of the meson, the quark and the antiquark.

In the spirit of the Brodsky-Lepage formalism (see also Ref.~\cite{bena}),
we define the four-vectors $q^{\mu}$ and $\bar{q}^{\mu}$ of the
quark and antiquark, respectively, as follows~:

\begin{eqnarray} 
 q^{\mu}=xQ^{\mu}+k^{\mu}~~~~~~~~~~~~~~~~~~~~~~~~
 { \bar{q}}^{\mu}=(1-x)Q^{\mu} - k^{\mu}\; ,
 \label{qmu}
\end{eqnarray}

\noindent
where $Q^{\mu}$ is the meson four-momentum, and $k^{\mu}$ is
the four-dimensional generalization of the above-mentioned
Fermi momentum (in the meson rest frame its energy and momentum
components are~: 0, $\bfk$). 
The energy of the quark resp. antiquark in the meson rest frame,
as derived from Eq.~(\ref{qmu}), is given by~:

\begin{eqnarray} 
 q_0=\sqrt{m^2+\bfk^2}=xM~~~~~~~~~~~
 \bar q_0=\sqrt{\bar m^2+\bfk^2}=(1-x)M\; .
 \label{q0}
\end{eqnarray}

Eq.~(\ref{mabqc}) is thus generalized, in the spirit of
Brodsky-Lepage, into~: 

\begin{eqnarray}
 {\cal M}  & = & \frac{1}{\sqrt{2M}}
 \int \frac{d^{3}\bfk}{(2 \pi)^{3/2}} 
 \Psi^{\textstyle *}(\bfk) \int
 \frac{ \Phi^{\textstyle *}_{N}(x) dx}{\sqrt{x(1-x)}}\,
 {\cal T}\; ,
 \label{mgene}
\end{eqnarray}

\noindent
where $\Phi_{N}(x)$ is the ($q\,\bar{q}$) meson's distribution
amplitude, normalized so that $\int \Phi_{N}(x) dx = 1 $.

To be precise, we notice that the amplitude ${\cal M}$
is a function of $E$ (the total c.m. energy of the process)
and of the scattering angle $\Theta$ [see Fig. 1(A)], while   
${\cal T}$ is a function of
$E,~\Theta,~k\,(\equiv |\bfk|)$ and of the angles $\theta$ and 
$\phi$ [see Fig. 1(B)] as well as of $x$.
On the other hand ${\cal M}$
depends on the following quantum numbers~:
$L$ (the meson's internal orbital angular momentum),
$S$ (its intrinsic spin), $\Lambda_{L}$ (the component of $L$
 on the $z$-axis of Fig.~1), $\Lambda_{S}$ (the component of
$S$ on that axis), and in addition on the helicities
$\lambda_{a},~\lambda_{b},~\lambda_{c}$ of particles $a$, $b$,
$c$ respectively~; ${\cal T}$
depends on $S,~\Lambda_{S}$ and
$\lambda_{a},~\lambda_{b},~\lambda_{c} $.

We then apply the relation~:

\begin{eqnarray}
 {\cal M}^{LSJ \Lambda}_{\lambda_{a} \lambda_{b} \lambda_{c} } 
 (E, \Theta) & = & \sum_{ \Lambda_{L}}\,
 \langle L S \Lambda_{L} \Lambda_{S} ~ |~ LSJ \Lambda \rangle\,
 {\cal M}^{LS \Lambda_{L} \Lambda_{S}}_{\lambda_{a} \lambda_{b}
 \lambda_{c}}(E, \Theta)\; ,
 \label{mlsj}
\end{eqnarray}

\noindent
where $J$ is the meson's total spin and $\Lambda$
(= $\Lambda_{L} + \Lambda_{S}$) the corresponding component
on the $z$-axis, and where we use the conventional notation  
$\langle j_{1} j_{2} m_{1} m_{2}~ |~ j_{1} j_{2} j m \rangle$
for Clebsch-Gordan coefficients~; on the other hand we factorize
the meson wave function into a radial and an angular part,
as follows~: 

\begin{eqnarray} 
 \Psi(\bfk) = R_{L}(k)~ Y_{L \Lambda_{L}} ( \theta, \phi)\; ,
 \label{psik}
\end{eqnarray}

\noindent
and finally, using the well-known expression
of the spherical harmonics~:

\begin{eqnarray} 
 Y_{L \Lambda_{L}} ( \theta, \phi)  & = &
 \sqrt{ \frac{2 L + 1}{4 \pi}}~ 
 d^{L}_{ \Lambda_{L} 0} ( \theta)~ e^{ i \Lambda_{L} \phi}\; ,
 \label{ylm}
\end{eqnarray} 

\noindent
where $ d^{L}_{ \Lambda_{L} 0} (\theta) $ is a Wigner rotation
matrix element, Eq.~(\ref{mgene}) becomes~:

\begin{eqnarray}
 {\cal M}^{LS J \Lambda}_{\lambda_{a} \lambda_{b} \lambda_{c}}
 (E, \Theta) & = & \frac{\sqrt{ 2 (2 L + 1) \pi}} {\sqrt{M}} 
 \int \frac{k^{2} dk}{(2 \pi)^{3/2}} ~ R_{L}^{\textstyle *}(k) 
 \int \frac{d(\cos\theta)}{2} \int \frac{d \phi}{2 \pi}
 \nonumber \\ & \times & \sum_{\Lambda_{L}}
 \zeta^{LSJ \Lambda_{L} \Lambda_{S}}( \theta, \phi )
 \int \frac{\Phi^{\textstyle *}_{N}(x) dx}{\sqrt{x(1-x)}} 
 {\cal T}^{S\,\Lambda_{S}}_{\lambda_{a}\lambda_{b}\lambda_{c}} 
 (E, \Theta, k, \theta, \phi , x) \; ,
 \label{mbl1}
\end{eqnarray} 

\noindent
where we define~:  

\begin{eqnarray} 
 \zeta^{LSJ \Lambda_{L} \Lambda_{S}}( \theta, \phi) & = &
 \langle LS \Lambda_{L} \Lambda_{S} ~|~ LSJ \Lambda \rangle \,
 d^{L}_{ \Lambda_{L} 0} ( \theta)~ e^{ - i \Lambda_{L} \phi}\; .
 \label{zeta}
\end{eqnarray}

Let us now assume that, in the c.m. frame of the reaction
$ ab \to Q c $, the meson is extreme-relativistic, i.e.
$M \ll E$~; in other words, the c.m. frame is to be considered
an ``infinite-momentum frame''. It is then easily shown (see 
Appendix A) that ${\cal T}$ becomes independent of $\phi$, so that,
because of the exponential in Eq.~(\ref{zeta}), only contributions
with $\Lambda_{L}=0$ (and consequently $ \Lambda = \Lambda_{S}$)
remain finite \footnote{\,
 $\Lambda$ is thus limited to the values $0,~ \pm 1$. It results
 (as already noticed in Ref.~\cite{bena}) that, when a prompt meson
 is found in a helicity state $\pm 2$ in a high-energy reaction,
 there is a good chance that it might be a glueball.}.
Eq.~(\ref{mbl1}) is thus simplified into~:

\begin{eqnarray}
 {\cal M}^{LS J \Lambda}_{\lambda_{a} \lambda_{b} \lambda_{c}}
 (E, \Theta) & = & 
 \frac{\sqrt{ 2(2 L + 1) \pi}} {\sqrt{M}} 
 \int \frac{k^{2} dk}{(2 \pi)^{3/2}} ~ R_{L}^{\textstyle *}(k) 
 \int  \frac{d (\cos \theta )}{2} \nonumber \\ & \times & 
 \zeta^{LSJ \Lambda} ( \theta ) \int
 \frac{\Phi_{N}^{\textstyle *}(x) dx}{\sqrt{x(1-x)}}\,
 {\cal T}^{S\,\Lambda}_{\lambda_{a} \lambda_{b} \lambda_{c}} 
 (E, \Theta, k, \theta, x)\; ,
 \label{mrel}
\end{eqnarray}

\noindent where 

\begin{eqnarray} 
 \zeta^{LSJ \Lambda} ( \theta) & = &
 \langle LS0 \Lambda ~|~ LSJ \Lambda \rangle \,
 d^{L}_{0\,0} ( \theta)\; .
 \label{z00}
\end{eqnarray}

At this point we notice that expanding the integral over
$x$ and $\theta$, in the r.h. member of Eq.~(\ref{mrel}),
in increasing powers of $k$, we expect the resulting power
series to start with the term in $k^{L}$, i.e.~: 

\begin{eqnarray}
 \int&&\!\!\!\!\!\!\!\!
 \frac{d( \cos \theta)}{2} \zeta^{LSJ \Lambda} (\theta) 
 \int \frac{ \Phi_{N}^{\textstyle *}(x) dx}{\sqrt{x(1-x)}}\,
 {\cal T}^{S\,\Lambda}_{\lambda_{a} \lambda_{b} \lambda_{c}}
 (E, \Theta,k, \theta,x) \qquad\qquad\qquad\qquad
 \nonumber \\ &&\:= a_{L} k^{L} + a_{L+1} k^{L+1}
 + a_{L+2} k^{L+2} + \dots \;\: .
 \label{kexp}
\end{eqnarray}

We now assume, as in \cite{cahn}, that in the meson rest frame
the quarks are essentially nonrelativistic, i.e. their
Fermi-momentum distribution is sharply peaked towards
$ k \to 0 $, so that in the r.h. member of Eq.~(\ref{kexp})
only the lowest-order term is to be retained. We then set~: 

\begin{eqnarray}
 \int \frac{k^{2} dk }{(2 \pi)^{3/2}}\, k^{L}
 R_{L}(k) & = & C_{L}\; .
 \label{cl1}
\end{eqnarray}

As has been shown in Refs.~\cite{cahn,kada}, this parameter
$C_{L}$ can be connected as follows with the meson wave function
at the origin in configuration space~: 

\begin{eqnarray}
 C_{L} & = & \frac{ (-i)^{L} (2 L + 1 )!!}{4 \pi L!} 
 \biggl[ \frac{d^{L}}{dr^{L}} R_{L}(r) \biggr]_{r \to\, 0}\; .
 \label{cl2}
\end{eqnarray}

Replacing the variable $k$ by the dimensionless one
$ \beta = 2k/M $, and setting 

\begin{eqnarray}
 f_{L} & = & \biggl(\frac{2}{M} \biggr)^{L+1/2}
 \sqrt{(2L+1) \pi}~C_{L}\; ,
 \label{fl}
\end{eqnarray}

\noindent
we now rewrite Eq.~(\ref{mrel}) as follows~: 

\begin{eqnarray}
 {\cal M}^{LS J \Lambda}_{\lambda_{a} \lambda_{b} \lambda_{c}}
 (E, \Theta) \!\!& = & \!\!f_{L}^{\textstyle *} \lim_{\beta \to 0}
 \frac{1}{\beta^{L}} \int \frac{ d ( \cos \theta )}{2}  
 \zeta^{LSJ \Lambda} ( \theta) \int
 \frac{\Phi_{N}^{\textstyle *}(x) dx}{\sqrt{x(1-x)}} 
 {\cal T}^{S\,\Lambda}_{\lambda_{a} \lambda_{b} \lambda_{c}} 
 (E, \Theta, \beta, \theta, x)\; . \nonumber \\
 \label{mfin}
\end{eqnarray}  

It can easily be checked that, for $L=0$, Eq.~(\ref{mfin})
leads exactly to the same expression as provided by the
Brodsky-Lepage formalism, defining the distribution amplitude
$\Phi(x)$ used by those authors as~: $\Phi(x) = f_0 \Phi_{N}(x)$.
Using the charged meson's leptonic decay for normalization, one
gets~: $|f_0| = f_M/(2\sqrt{3})$, where $f_M$ is the meson's
leptonic decay constant~; in particular, $f_\pi\cong 93$ MeV.

Eq.~(\ref{mfin}) is easily generalized for the case of
($q\,{\bar q}$) meson pair production, i.e. $ab \to QQ'$~;
one gets

\begin{eqnarray}
 {\cal M}^{LS J \Lambda,~L'S'J'\Lambda'}_{\lambda_{a} \lambda_{b}} 
 (E, \Theta) \!\!\!& = &\!\!\! 
 f_L^{\textstyle *}~f_L^{\prime \textstyle{*}} \lim_{\beta,~\beta' \to 0}
 \frac{1}{\beta^{L} \beta^{\prime L'}} 
 \int \frac{d ( \cos \theta )}{2} \zeta^{LSJ \Lambda} ( \theta)   
 \int \frac{ d (\cos \theta')}{2} \zeta^{L'S'J' \Lambda'} ( \theta') 
 \nonumber \\ &\!\!\! \times \!\!\!& \!\!\!\!
 \int \frac{\Phi_{N}^{\textstyle *}(x) dx}{\sqrt{x(1-x)}}
 \int \frac{\Phi_{N}^{\prime\textstyle *}(x') dx'}{\sqrt{x'(1-x')}}\,
 {\cal T}^{S\,\Lambda,\,S'\,\Lambda'}_{\lambda_{a} \lambda_{b}} 
 (E, \Theta, \beta, \beta', \theta, \theta', x , x')\; ,
 \nonumber \\
 \label{mqq}
\end{eqnarray}   

\noindent
where ${\cal M}$ refers to the process $ ab \to QQ' $,
while ${\cal T}$ refers to $ ab \to q {\bar q} q' {\bar q}'$
and where all symbols without and with ``prime'' are 
pertaining to $Q$ and $Q'$ respectively.

The expressions of momentum four-vectors and projectors of
spinor pairs, to be used in the calculation of the amplitudes
${\cal T}$, both for $ab \to Qc$ and $ab \to Q Q'$, are shown
in Appendix A.    

\goodbreak

\section{Application to the production of tensor and
 pseudotensor mesons in photon-photon collisions}

\nobreak

We shall here consider the following processes~:

\begin{enumerate}
\item[{\bf\it (i)}] Pair production, in $\gamma\gamma$ collisions,
 of $(q\,\bar q)$ tensor mesons, i.e. of $f_2(1270)$,
 $a_2(1310)$ and $f'_2(1525)$, all of them with quantum
 numbers $L = S = 1$, $J=2$.
\item[{\bf\it (ii)}] Pair production, in $\gamma\gamma$ collisions,
 of $(q\,\bar q)$ pseudotensor mesons $\pi_2(1670)$ with
 $L=2$, $S=0$, $J=2$.
\item[{\bf\it (iii)}] Production, in $\gamma\gamma$ collisions,
 of hybrid pairs made of one pion ($L=S=J=0$) and one
 $\pi_2$ meson ($L=2,S=0,J=2$).
\end{enumerate}

For those three types of reactions, Eq.~(\ref{mqq})
becomes respectively~:

\begin{eqnarray}
% \mbox{\bf (i)}\>
 {\cal M}^{1\:\!1\:\!2\:\!\Lambda,\:\!1\:\!1\:\!2\:\!\Lambda'}
 _{\lambda_{\gamma} \lambda'_{\gamma}} (E, \Theta) 
 & = & f_1^{\textstyle *}\,f^{\prime \textstyle{*}}_1\,
 \lim_{ \beta,\,\beta' \to 0 } 
 \frac{1} {\beta \beta'} 
 \int \frac{d(\cos \theta)}{2}\ 
 \zeta^{1\:\!1\:\!2\:\!\Lambda}( \theta )   
 \int \frac{d(\cos \theta')}{2}\ 
 \zeta^{1\:\!1\:\!2\:\!\Lambda'}( \theta' )   
 \nonumber\\ & \times &
 \int \frac{\Phi_{N}^{\textstyle *}(x)\,dx}{\sqrt{x(1-x)}}
 \int \frac{\Phi^{\prime \textstyle{*}}_{N}(x')\,dx'}{\sqrt{x'(1-x')}}
 \,{\cal T}^{ 1\:\!\Lambda,\:\!1\:\!\Lambda'}
 _{\lambda_{\gamma} \lambda_{\gamma}'} 
 (E, \Theta, \beta,\beta', \theta , \theta', x, x') \; ,
 \nonumber \\
 & &
 \label{mtt}
\end{eqnarray}

\begin{eqnarray}
% \mbox{\bf (ii)}\>\;
 {\cal M}^{2\:\!0\:\!2\:\!0,\:\!2\:\!0\:\!2\:\!0}
 _{\lambda_{\gamma} \lambda'_{\gamma}} (E, \Theta) 
 & = & f_2^{\textstyle *}\,f^{\prime \textstyle{*}}_2\,
 \lim_{ \beta,\,\beta' \to 0 } 
 \frac{1} {\beta^2 \beta'^2} 
 \int \frac{d(\cos \theta)}{2}\ \zeta^{2\:\!0\:\!2\:\!0}( \theta )   
 \int \frac{d(\cos \theta')}{2}\ \zeta^{2\:\!0\:\!2\:\!0}( \theta' )   
 \nonumber\\ & \times &
 \int \frac{\Phi_{N}^{\textstyle *}(x)\,dx}{\sqrt{x(1-x)}}
 \int \frac{\Phi^{\prime \textstyle{*}}_{N}(x')\,dx'}{\sqrt{x'(1-x')}}
 \,{\cal T}^{0\:\!0,\:\!0\:\!0}
 _{\lambda_{\gamma} \lambda_{\gamma}'} 
 (E, \Theta, \beta,\beta', \theta , \theta', x, x') \; ,
 \label{mps}
\end{eqnarray}

\begin{eqnarray}
% \mbox{\bf (iii)}\>\>\>
 {\cal M}^{0\:\!0\:\!0\:\!0,\:\!2\:\!0\:\!2\:\!0}
 _{\lambda_{\gamma} \lambda'_{\gamma}} (E, \Theta) 
 & = & f_0^{\textstyle *}\,f^{\prime \textstyle{*}}_2\,
 \lim_{ \beta' \to 0 } 
 \frac{1} { \beta'^2} 
 \int \frac{d(\cos \theta')}{2}\ \zeta^{2\:\!0\:\!2\:\!0}( \theta' )   
 \nonumber\\ & \times &
 \int \frac{\Phi_{N}^{\textstyle *}(x)\,dx}{\sqrt{x(1-x)}}
 \int \frac{\Phi^{\prime \textstyle{*}}_{N}(x')\,dx'}{\sqrt{x'(1-x')}}
 \,{\cal T}^{0\:\!0,\:\!0\:\!0}
 _{\lambda_{\gamma} \lambda_{\gamma}'} 
 (E, \Theta,\beta',\theta', x, x') \; .
 \label{mpi}
\end{eqnarray}

In writing down Eq.~(\ref{mpi}) we have made use of the fact
that, for $\beta \to 0$, the amplitude
${\cal T}^{0\:\!0,\:\!0\:\!0}
_{\lambda_{\gamma} \lambda_{\gamma}'}$ here becomes independent
of $\theta$ (as can easily be checked), and that in addition
one has~: $\int \left[d(\cos\theta)/2\right]
\zeta^{0\:\!0\:\!0\:\!0}(\theta)=1$.

In order to use Eqs.~(\ref{mtt})--(\ref{mpi}) for the
evaluation of the amplitudes at the hadron level, we
must choose the distribution amplitudes $\Phi_N(x)$,
$\Phi'_N(x')$ and, on the other hand, determine the values
of the constants $f_L$, $f'_{L'}$ involved.

For the distribution amplitudes (DA) of the tensor and
pseudotensor mesons, we shall make two different choices~:

\begin{description}
\item[{\it (i)}] The so-called nonrelativistic DA \cite{brle}~:
 \begin{equation}
  \Phi_N(x) = \delta(x-\frac{1}{2})\; .
  \label{danr}
 \end{equation}
\item[{\it (ii)}] A generalization of the so-called asymptotic
 DA \cite{brle}, namely~:
 \begin{equation}
  \Phi_N(x) = \frac{2^{2L+3}\,\Gamma(L+5/2)}
  {\sqrt{\pi}\,\Gamma(L+2)}\, x^{L+1}(1-x)^{L+1} \; ,
  \label{dasy}
 \end{equation}
\end{description}

\noindent
%i.e. more precisely
where $\Gamma(z)$ is the well-known Gamma function; i.e.

\begin{equation}
 \Phi_N(x) = 30x^2(1-x)^2
 \qquad\qquad \mbox{\rm and} \qquad\qquad
 \Phi_N(x) = 140x^3(1-x)^3
 \label{da23} 
\end{equation}

\noindent
for tensor mesons and pseudotensor mesons respectively
(notice that, for $L \neq 0$, using the asymptotic DA
without modification, i.e. Eq.~(\ref{dasy}) for $L=0$,
would lead to divergences).

Finally, for the pion, we shall use either the asymptotic
DA \cite{brle}~:

\begin{equation}
 \Phi_N(x) = 6x(1-x)\; ,
 \label{asypi}
\end{equation}

\noindent
or the Chernyak-Zhitnitsky DA \cite{cher}
\footnote{\,We are aware that both of those pion DA's are partly
unsatisfactory as regards practical applications~:
the asymptotic DA does not fit the pion's electromagnetic
form factor, while the Chernyak-Zhitnitsky DA is disfavoured
because of its incompatibility with the photon-pion
transition form factor. We are using them nevertheless, as we
did in previous papers \cite{icho,murg}, since presently
no ``gold-plated'' pion DA is available.}~:

\begin{equation}
 \Phi_N(x) = 30x(1-x)(1-2x)^2 \; .
 \label{dacz}
\end{equation}

As for the constants $f_L$, we extract them (more precisely~:
their absolute values) from the known experimental values
of the decay widths $\Gamma(Q\to\gamma\gamma)$, as we
compute the corresponding theoretical expressions by
applying again the formalism presented in Sect. 2
(see Appendix C).

There is then no free parameter left in the calculation
of the amplitudes ${\cal M}$ of the various processes
considered, except for an irrelevant phase factor and
for some limited freedom in the choice of the value of
the strong coupling constant.

{}From those amplitudes the differential cross section
with respect to $\cos\Theta$ is derived as follows
(neglecting meson masses, as well, in the phase-space
factor)~:

\begin{eqnarray}
 \frac{d \sigma^{ \gamma \gamma' \to QQ'}(E,\Theta)}
 {d ( \cos \Theta)} & = & \frac{\xi}{128 \pi E^{2}}
 \sum_{\lambda_{\gamma} \lambda_{\gamma}',~\Lambda \Lambda'} 
 |~ {\cal M}^{LSJ\Lambda,L'S'J' \Lambda'}
 _{\lambda_{\gamma} \lambda_{\gamma}'}(E, \Theta)|^{2}\; ,
 \label{dsdt}
\end{eqnarray}

\noindent
where the coefficient $\xi$ takes the value 1/2 when $ Q,~Q' $
are identical particles, and 1 otherwise. 

Details of our calculations, for the three processes
$\gamma\gamma\to Q Q'$ considered, are shown in Appendix B.
Only for the choice of the nonrelativistic DA are those
calculations performed analytically till the end.

Choosing the value $\alpha_{s}=0.3$, we are plotting in
Figs. 2 and 3, for the three types of reactions considered
and for the known charged or neutral tensor and pseudotensor
mesons to be produced in these reactions, the scaling 
differential cross section $ E^{8} d \sigma /d|t|$,
which is easily derived  from Eq.~(\ref{dsdt}) [noticing
that, neglecting masses, $ |t|=(E^{2}/2)(1 - \cos \Theta)$].
Figs. 2(a),(b) correspond to the case of tensor-meson
pairs, making use of the nonrelativistic DA [Eq.~(\ref{danr})]
and of the generalized asymptotic DA [Eq.~(\ref{da23})]
respectively. Similarly, Figs. 3(a),(b) show the case of
pseudotensor-meson pairs and of hybrid (pion plus
pseudotensor meson) pairs.

On the other hand we are interested in the $p_T$ distribution
of the outgoing mesons. Replacing the variable $\cos\Theta$ by 
$p_{T}$, we get~:

\begin{eqnarray}
 \frac{d \sigma ^{\gamma \gamma' \to Q Q'}
 (E,\,p_{T})}{d p_{T}} & = & 
 \frac{p_{T}\,\xi}{64 \pi E^{3} \sqrt{p^{2}-p^{2}_{T}}} 
 \sum_{ \lambda_{\gamma} \lambda_{\gamma}' \Lambda \Lambda'} 
 \left[| {\cal M}_{\lambda_{\gamma} \lambda_{\gamma}'}
 ^{\Lambda \Lambda'}(E,\Theta)|^{2} + 
 | {\cal M}_{\lambda_{\gamma} \lambda_{\gamma}'}
 ^{\Lambda \Lambda'} (E, \pi-\Theta) |^{2}\right]\; , \nonumber \\
 &  &
 %\mbox{~~~~~~~~~~~~~~~~}
 \label{dsdpt}
\end{eqnarray}

\noindent 
where, in the expressions of the amplitudes, $\cos\Theta$ is
to be replaced by $[1- p^{2}_{T}/p^{2}]^{1/2}$, $p^{2}$ being
given by

\begin{equation}
 p^2 = \frac{1}{4E^2}\left[E^4-2\left(M^2+M'^2\right)E^2+
 \left(M^2-M^{\prime\,2}\right)^2\right]
 \label{p2}
\end{equation}

\noindent
(It is indeed preferable to keep the meson masses finite in the
expression of $p^2$, if one wishes to extrapolate the $p_T$
spectrum to relatively small values of $p_T$).

{}From Eq.~(\ref{dsdpt}) one derives the transverse-momentum
spectrum for the overall reaction $ee' \to ee'QQ' $~:    

\begin{eqnarray}
 \frac{d \sigma ^{e e' \to e e'Q Q'}
 (s,\,p_{T})}{d p_{T}} & = & 
 \int^{1}_{z_{min}} f(z) dz \int^{1}_{z'_{min}} f(z') dz' 
 \,\frac{d \sigma ^{\gamma \gamma' \to Q Q'}
 (E,~p_{T})}{d p_{T}}\; ,
 \label{dsee}
\end{eqnarray}

\noindent
where $s$ is the overall c.m. energy squared~; $ f(z) $ resp.
$ f(z') $, is the equivalent-photon spectrum of either
electron, given by 

\begin{eqnarray}
 f(z) = \frac{\alpha}{\pi z} \biggl[ (1-z+ \frac{z^{2}}{2})
 \ln \frac{s}{4 \rm{m}_{\rm{e}}^{2}} - (1-z) \biggr]\; ,
 \label{phot}
\end{eqnarray}

\noindent
and we notice that

\begin{equation}
 E^{2} = z z' s\;, \qquad z'_{min} = \frac{E_{min}^{2}}{zs}\;,
 \qquad z_{min} = \frac{E_{min}^{2}}{s}\;,
 \label{zz}
\end{equation}

\noindent where

\begin{equation}
 E_{min} = (p_{T}^{2} + M^{2})^{1/2} +
 (p_{T}^{2} + M'^{2})^{1/2}\; .
 \label{emin}
\end{equation}

As for the running strong-interaction coupling constant,
we here use the lowest-order (one-loop) formula~:

\begin{eqnarray}
 \alpha_{s} & = & \frac{12 \pi}
 {25 \ln (\mbox{``}Q^2\mbox{''}/ \Lambda^{2}_{\rm{QCD}})}\; ,
 \label{astrong}
\end{eqnarray}

\noindent
where an empirical value of 200 MeV is taken for $\Lambda_{QCD}$.
Regarding the scale ``$Q^2$'',
we associate it, in the spirit of Refs. \cite{mack,crji}, with the
mean absolute value (i.e., the absolute value for $x=x'=1/2$)
of the squared four-momentum of the timelike or spacelike gluon
exchanged in the Feynman diagrams for $\gamma\gamma\to
q\,\bar{q}\,q'\,\bar{q}'$; thus $\mbox{``}Q^2\mbox{''}\approx E^2/4$.

The numerical integration over $z$, $z'$ and $p_T$
leads to the values shown in tables 1-3 for the 
integrated cross sections, respectively for the three
cases~: {\it i)} $\sqrt{s}=200$ GeV (LEP2 energy), $p_T > 1$ GeV~;
{\it ii)} $\sqrt{s}=200$ GeV, $p_T > 2$ GeV~;
{\it iii)} $\sqrt{s}=10$ GeV (energy of a ``B factory''),
$p_T > 1$ GeV.

The three types of reactions, and both types of distribution
amplitudes here defined for tensor and pseudotensor mesons,
are considered.

\goodbreak

\section{Conclusion}

\nobreak

As is shown by tables 1-3 (and as appears already in Figs. 2,3),
the results obtained do not differ widely depending on the
distribution amplitude chosen for tensor resp. pseudotensor
mesons. Except for the case of $\pi_2^0\pi_2^0$ production,
the generalized asymptotic DA leads to approximately equal
or slightly (at most by a factor of about 3) higher values,
as compared with the nonrelativistic one.

On the other hand, comparing the Chernyak-Zhitnitsky
distribution amplitude with the asymptotic one for
the pion in the case of hybrid-pair production, one notices
that the former gives rise to somewhat higher yields
(by a factor of 2-4) than the latter.

Finally it should be remarked that in general the charged
channels give rise to significantly higher yields than the
neutral ones. There appears to be some hope that the production
of charged-meson pairs as here considered may become measurable
with high-energy $e^+e^-$ colliders of the next generation,
provided integrated luminosities as high as $\approx 10^{40}$
cm$^{-2}$ can be reached.

\goodbreak

\setcounter{equation}{0}
\renewcommand{\theequation}{\mbox{A}\arabic{equation}}
\section*{ Appendix A~: Expression of four-vectors and of
 spinor-pair projectors} 

\nobreak

For the four-momenta of initial and final particles,
resp. partons, involved in the process $ ab \to Q c $,
we get in the c.m. frame of that process [Fig. 1(A)],
assuming the masses of $a,~b,~c$ to be zero or negligible, the
following expressions (components $0,\,x,\,y,\,z$ in that order)~:

\begin{equation}
 a^{\mu}= \frac{E}{2} \left(\begin{array}{c}
 1 \\
 - \sin \Theta \\
 0 \\
{} ~ \cos \Theta
 \end{array} \right)~~~~~~~~~~~~~~~~~~~~
 b^{\mu}= \frac{E}{2} \left(\begin{array}{c}
 1 \\
 {}~ \sin \Theta \\
 0 \\
 - \cos \Theta
 \end{array} \right) \\  
 \label{amu}
\end{equation}

\begin{equation}
 Q^{\mu}= \frac{E}{2} \left(\begin{array}{c}
 1 + \eta^{2} \\
 0 \\
 0 \\
 1 - \eta^{2} 
 \end{array} \right)~~~~~~~~~~~~~~
 c^{\mu}= \frac{E}{2} (1 - \eta^{2} ) \left(\begin{array}{c}
 1 \\
 0 \\
 0 \\
 -1
 \end{array} \right)   \; ,
 \label{cmu}
\end{equation}

\noindent
where we have called all four-momenta like the corresponding
particles and set $\eta = M/E$.

For the four-momenta of the quarks $q, \bar{q}$ we get, after
a Lorentz transformation from the meson rest frame [Fig. 1(B)]
to the c.m. frame of the process considered [Fig. 1(A)]~:

\begin{equation}
 q^{\mu} = \frac{Ex}{2} \left(\begin{array}{c}
 1 + \beta_{q} \cos \theta +
 \eta^{2} (1 - \beta_{q} \cos \theta) \\
 2 \eta \beta_{q} \sin \theta \cos \phi \\
 2 \eta \beta_{q} \sin \theta \sin \phi \\
 1 + \beta_{q} \cos \theta -
 \eta^{2} (1 - \beta_{q} \cos \theta)  
 \end{array} \right)~~~
 {\bar q}^{\mu} = q^{\mu} (x \to 1-x,~\beta_{q}
 \to - \beta_{\bar q})\; , 
 \label{qcm}
\end{equation}

\noindent
where we have set~:

\begin{equation}
 \beta_{q} = \frac{k}{Mx} = \frac{\beta}{2x},\quad
 \beta_{\bar q }=\frac{k}{M(1-x)} = 
 \frac{\beta}{2(1-x)}\; .
 \label{bk}
\end{equation}

Letting $\eta$ go to zero, Eq.~(\ref{amu}) remains unmodified,
while Eqs.~(\ref{cmu}),(\ref{qcm}) are replaced by the
simplified formulas (\ref{cmusi}),(\ref{qcmsi}) as follows 

\begin{equation}
 Q^{\mu}= \frac{E}{2} \left(\begin{array}{c}
 1 \\
 0 \\
 0 \\
 1  
 \end{array} \right)~~~~~~~~~~~~~~
 c^{\mu}= \frac{E}{2} \left(\begin{array}{c}
 1 \\
 0 \\
 0 \\
 -1
 \end{array} \right)   
 \label{cmusi}
\end{equation}

\begin{equation}
 q^{\mu}= \frac{Ex}{2} (1+\beta_{q} \cos \theta) \left(\begin{array}{c}
 1 \\
 0 \\
 0 \\
 1  
 \end{array} \right)~~~~~~~~~~~~~~
 \bar{q}^{\mu}= \frac{E (1-x)}{2} (1- \beta_{\bar q} \cos \theta) 
 \left(\begin{array}{c}
 1 \\
 0 \\
 0 \\
 1
 \end{array} \right) \; .
 \label{qcmsi}
\end{equation}

{}From Eq.~(\ref{qcmsi}) one derives the
expressions of the projectors of spinor pairs to be used
in the computation of the helicity amplitudes at the parton level. 
We notice that, in the expressions of the quark four-vectors
(and consequently as well of the quark spinors), any
dependence on $\phi$ has vanished. 
Therefore, as stated in Sect. 2, all helicity amplitudes ${\cal T}$
become independent of $\phi$. Calling $u^{\lambda_{q}}$,
$v^{\lambda_{ \bar q}}$ the spinors of the quark and antiquark,
with helicity $\lambda_{q}$ and $\lambda_{ \bar q}$ respectively,
and introducing the Clebsch-Gordan coefficient 
$\langle\,\frac{ \scriptstyle 1}{ \scriptstyle 2} 
\frac{ \scriptstyle{1}}{ \scriptstyle{2}}
\lambda_{q} \lambda_{ \bar{q}} | 
\frac{ \scriptstyle 1}{ \scriptstyle 2} 
\frac{ \scriptstyle 1}{ \scriptstyle 2} S \Lambda\,\rangle$,
one gets the four projectors $P^{S}_{\Lambda}$ needed~:

\begin{eqnarray}
 P^{1}_{1} & = & v^{1/2} {\bar u}^{1/2} \nonumber \\
 & = &  \frac{1}{\sqrt{2}} \sqrt{x(1-x)} 
 \sqrt{(1+\beta_{q} \cos \theta)(1-\beta_{\bar{q}} \cos \theta)}~ 
 \slash{\epsilon^{\textstyle *}_{+}} \slash{Q} \; ,
 \label{pr11}
\end{eqnarray} 

\begin{eqnarray}
 P^{1}_{0} & = & \frac{1}{\sqrt{2}}
 (v^{-1/2} \bar{u}^{1/2}+ v^{1/2} \bar{u}^{-1/2}) \nonumber \\
 & = & \frac{1}{ \sqrt{2} } \sqrt{x(1-x)}
 \sqrt{(1+\beta_{q} \cos \theta)(1-\beta_{\bar{q}} \cos \theta)}
 \slash{Q} \; ,
 \label{pr10}
\end{eqnarray}

\begin{eqnarray}
 P^{1}_{-1} & = & \frac{1}{\sqrt{2}}  v^{-1/2}
 \bar{u}^{-1/2} \nonumber \\ 
 & = & \frac{1}{ \sqrt{2}} \sqrt{x(1-x)}
 \sqrt{(1+\beta_{q} \cos \theta)(1-\beta_{\bar{q}} \cos \theta)} 
 \slash{\epsilon^{\textstyle *}_{-}} \slash{Q}  \; ,
 \label{pr1m1}
\end{eqnarray} 

\begin{eqnarray}
 P^{0}_{0} & = & \frac{1}{\sqrt{2}}
 (v^{-1/2} \bar{u}^{1/2} - v^{1/2} \bar{u}^{-1/2}) \nonumber \\ 
 & = & \frac{1}{\sqrt{2}} \sqrt{x(1-x)}
 \sqrt{(1+\beta_{q} \cos \theta)(1-\beta_{\bar{q}}
 \cos \theta)} \gamma_{5} \slash{Q} \; .
 \label{pr00}
\end{eqnarray}

Here the four-vectors $ \epsilon^{\textstyle *}_{+}$,
$\epsilon^{\textstyle *}_{-}$ are defined in the usual way~: 

\begin{eqnarray}
 \epsilon^{{\textstyle *} \mu}_{+} = - \frac{ \displaystyle 1}
 { \displaystyle \sqrt{2} }
 \left(\begin{array}{c}
 0\\
 1\\
 -i\\
 0
 \end{array} \right)~~~~~~~~~~~~~~~~~~~~~~~~~~
 \epsilon^{{\textstyle *} \mu}_{-} = \frac{ \displaystyle 1}
 { \displaystyle \sqrt{2} }
 \left(\begin{array}{c}
 0\\
 1\\
 i\\
 0
 \end{array} \right) \; .
 \label{pol}
\end{eqnarray}

When one goes over from the process $ab \to Qc$ to 
$ab \to QQ'$, $c^{\mu}$ is replaced by $Q^{\prime \mu}$
in Eq.~(\ref{cmusi}), and in addition the following expressions
are to be used for the four-vectors \mbox{$q'^{\mu},
{}~{\bar q}'^{\mu}$~:} 

\begin{eqnarray}
 q'^{\mu} =  \frac{ \displaystyle Ex'}
 { \displaystyle 2 } (1 - \beta'_{q'} 
 \cos \theta')
 \left(\begin{array}{c}
 1\\
 0\\
 0\\
 -1
 \end{array} \right)~~~~~~~
 {\bar q}'^{\mu} =  \frac{ \displaystyle E (1-x')}
 { \displaystyle \sqrt{2} }
 (1+ \beta'_{\bar{q}'} \cos \theta')
 \left(\begin{array}{c}
 1\\
 0\\
 0\\
 -1
 \end{array} \right) \; ,
 \label{qpmu}
\end{eqnarray}

\noindent
defining

\begin{eqnarray} 
 \beta'_{q'} = \frac{ \displaystyle \beta' }
 { \displaystyle 2x'},~~ 
 \beta'_{\bar{q}'} = \frac{ \displaystyle \beta' }
 { \displaystyle 2(1-x')}  \; .
 \label{beta}
\end{eqnarray}

As for the corresponding projectors of spinor pairs,
$P^{\prime S'}_{\Lambda'}$, they are written as follows~:

\begin{eqnarray}
 P'^{1}_{1} & = & 
 \frac{1}{\sqrt{2}} \sqrt{x'(1-x')} 
 \sqrt{(1 - \beta'_{q'} \cos \theta')
 (1 + \beta'_{\bar{q}'} \cos \theta')} 
 \slash{\epsilon'^{\textstyle{*}}_{+}} \slash{Q'} \; ,\\ 
 P'^{1}_{0} & = &  \frac{1}{ \sqrt{2} } \sqrt{x'(1-x')}
 \sqrt{(1 - \beta'_{q'} \cos \theta')
 (1 + \beta'_{\bar{q}'} \cos \theta)'} 
 \slash{Q'} \; ,\\
 P'^{1}_{-1} & = & \frac{1}{ \sqrt{2}} \sqrt{x'(1-x')}
 \sqrt{(1 - \beta'_{q} \cos \theta')
 (1 + \beta'_{\bar{q}'} \cos \theta')} 
 \slash{\epsilon'^{\textstyle{*}}_{-}} \slash{Q'} \; ,\\ 
 P'^{0}_{0} & = & \frac{1}{\sqrt{2}} \sqrt{x'(1-x')}
 \sqrt{(1 - \beta'_{q'} \cos \theta')
 (1 + \beta'_{\bar{q}'} \cos \theta)'} 
 \gamma_{5} \slash{Q'} \; ,
 \label{prqp}
\end{eqnarray}

\noindent
where one defines~:
$\epsilon'^{{\textstyle *} \mu}_{\pm} =
\epsilon^{{\textstyle *} \mu}_{\mp}$.

\goodbreak

\setcounter{equation}{0}
\renewcommand{\theequation}{\mbox{B}\arabic{equation}}
\section*{ Appendix B~: Details of calculation
 of the processes considered }

\nobreak

Using the ingredients given in Appendix A, we shall here
provide details of calculation of the three processes
considered in Sect. 3~; that calculation is based on
computing the usual Feynman diagrams involved in the
Brodsky-Lepage mechanism \cite{brle} for $\gamma\gamma$
production of $(q\,\bar q)$ meson pairs (e.g. $\gamma\gamma
\to\pi\pi)$. We shall give hereafter

\begin{enumerate}
\item[--] except for the case of hybrid-pair production
 (where they become too lengthy), the general expression
 of the physically relevant pieces of the
 helicity amplitudes ${\cal T}^{\Lambda\Lambda'}
 _{\lambda_\gamma\lambda'_\gamma}$ (dropping the superscripts
$S$, $S'$ of those amplitudes)~;
\item[--] the expressions of the amplitudes
 ${\cal M}^{\Lambda\Lambda'}_{\lambda_\gamma\lambda'_\gamma}$
 (dropping the superscripts $S,L,J,S',L',J'$), but only for
 the choice of the nonrelativistic distribution amplitude
 given by Eq.~(\ref{danr}) for $(q\,\bar q)$ tensor and
 pseudotensor mesons~; indeed, using the DA given by
 Eq.~(\ref{da23}), it becomes difficult, if not impossible,
 to perform the integration over both $x$ and $x'$, involved
 in Eqs.~(\ref{mtt})--(\ref{mpi}), analytically~;
\item[--] the expressions of the differential cross sections
 $d\sigma^{\gamma\gamma'\to QQ'}/d(\cos\Theta)$, again only
 for the choice of the nonrelativistic distribution amplitude
 for tensor and pseudotensor mesons.
\end{enumerate}

\noindent
{\bf{\it (i)} $\mbox{\boldmath $\gamma\gamma$}$ production of
 $\mbox{\boldmath $(q\,\bar q)$}$ tensor-meson pairs}

\vspace{6pt}

Keeping only the physically relevant pieces (i.e. the
terms in $\beta^1\beta^{\prime 1}$ of the series expansion in powers
of $\beta$, $\beta'$) of the helicity amplitudes
${\cal T}^{\Lambda\Lambda'}_{\lambda_\gamma\lambda'_\gamma}$,
we obtain~:

\begin{eqnarray}
 \left({\cal T}^{+\:\!-}_{\pm\:\!\mp}\right)_{phys.}
 & = & -\,\frac{K}{4(a^2-b^2)^2}\,
 \biggl\{\,\langle\,(e_{1}-e_{2})^{2}\rangle\,(2a-1)
 \nonumber \\ & \pm & 2\,v\,(1\mp v)
 \frac{\langle e_{1}e_{2}\rangle}
 {\left[a^{2}-v^{2}b^{2}\right]^{3}}\,
 \Bigr[\,a^2(2a^4-9a^2b^2+6a^3b^2+6b^4-4ab^4) \nonumber \\
 & + & 2v^2b^2(3a^4-3a^2b^2-4a^3b^2+b^4+2ab^4) +
 v^4(2a-1)b^6\,\Bigr]\biggr\}\,uu'\beta\beta' \; ,\nonumber \\
 \label{ttpmpm}
\end{eqnarray}

\begin{equation}
 \left({\cal T}^{-\:\!+}_{\mp\:\!\pm}\right)_{phys.} =
 \left({\cal T}^{+\:\!-}_{\pm\:\!\mp}\right)_{phys.}\; ,
 \qquad\qquad\qquad\qquad\qquad\qquad
 \label{ttmppm}
\end{equation}

\begin{eqnarray}
 \left({\cal T}^{0\:\!0}_{\pm\:\!\pm}\right)_{phys.}
 & = &
 -\,\frac{K}{2(a^2-b^2)^2}\,
 \frac{\langle\,(e_1-e_2)^2\rangle}{1-v^2}\,a\: uu'\beta\beta'
 \; , \qquad\qquad\qquad
 \label{ttpp}
\end{eqnarray}

\begin{eqnarray}
 \left({\cal T}^{0\:\!0}_{\pm\:\!\mp}\right)_{phys.} & = &
 \frac{K}{2(a^2-b^2)^2}\,
 \biggl\{\,\frac{\langle\,(e_1-e_2)^2\rangle}{1-v^2}\,(1-a)
 \qquad\qquad \nonumber \\
 & + & \frac{\langle e_1e_2\rangle}
 {\left[a^2-v^2b^2\right]^3}\,\Bigl[\,a^3(a^3-3a^2b^2+2b^4)
 \nonumber \\ &+& 2v^2a(-a^5+3a^3b^2+3a^4b^2-3ab^4-5a^2b^4
 +3b^6) \nonumber \\ &+&
 v^4b^2(-6a^4+9a^2b^2+2a^3b^2-2b^4-3ab^4)\,\Bigr]\biggr\}
 \,uu'\beta\beta' \; ,
 \label{ttpm}
\end{eqnarray}

\noindent
all other relevant helicity amplitudes  
${\cal T}^{\Lambda\Lambda'}_{\lambda_\gamma\lambda'_\gamma}$
being zero. In the above equations we have introduced the constant
$K=2^{10}\pi^2\alpha\alpha_s/(3E^2)$, using also for shorthand
$v=\cos\Theta$, $u=\cos\theta$, $u'=\cos\theta'$.
In addition we have put

\begin{eqnarray}
 a &=& (1-x)(1-x')+xx' \; ,\nonumber \\
 b &=& (1-x)(1-x')-xx' \; .
 \label{ab}
\end{eqnarray}

In standard notation,
$e_1$ and $-e_2$ are the charges, in units of the proton
charge, of the quark and the antiquark of which a given meson is
composed~; $\langle e_1e_2\rangle$ and
$\langle\,(e_1-e_2)^2\rangle$ are then charge factors, adequately
convoluted with the flavor components of the $Q$, $Q'$ meson
wave functions.
The numerical values of these charge factors for the various
meson pairs here considered are given in table B1.

Therefrom we get through Eq.~(\ref{mtt}), with the choice
$\Phi_N(x)=\delta(x-1/2)$, $\Phi'_N(x')=\delta(x'-1/2)$~:

\begin{equation}
 {\cal M}^{+\:\!-}_{\pm\:\!\mp} = \mp\ \frac{8}{9}\, K f_1 f_1'
 \,\langle e_1e_2 \rangle\,v(1 \mp v) \; ,
 \label{mttpmpm}
\end{equation}

\begin{equation}
 {\cal M}^{-\:\!+}_{\mp\:\!\pm} = {\cal M}^{+\:\!-}_{\pm\:\!\mp} \; ,
 \label{mttmppm}
\end{equation}

\begin{equation}
 {\cal M}^{0\:\!0}_{\pm\:\!\pm} = -\,\frac{8}{27}\, K f_1 f_1'
 \ \frac{\langle\,(e_1-e_2)^2\rangle}{1-v^2} \; ,
 \label{mttzpp}
\end{equation}

\begin{equation}
 {\cal M}^{0\:\!0}_{\pm\:\!\mp} = \frac{8}{27}\, K f_1 f_1'
 \left[\ \frac{\langle\,(e_1-e_2)^2\rangle}{1-v^2}
 +2\ \langle e_1e_2 \rangle\ (1-2v^2)\ \right] \; .
 \label{mttzpm}
\end{equation}

 Finally, with the choice of the nonrelativistic
DA for the tensor mesons, Eq.~(\ref{dsdt})
leads us to~:

\begin{eqnarray}
 \frac{d\sigma^{\gamma\gamma\to QQ'}}{d(\cos\Theta)} & = &
 \frac{2^{21}\pi^{3}\alpha^{2}\alpha_{s}^{2} |f_{1}|^{2}
 |f_{1}'|^{2}\,\xi}{3^{8}s^{3}}\,\biggl\{\,
 \frac{\,\langle\,(e_1-e_2)^2\rangle^2}{(1-v^2)^2} \nonumber \\
 & + & 2\,\langle e_1e_2\rangle\langle\,(e_1-e_2)^2\rangle\,
 \frac{1-2v^2}{1-v^2} + \langle e_1e_2\rangle^2
 (2+v^{2}+17v^{4})\biggr\} \; .
 \label{dsgqtt}
\end{eqnarray}

\noindent
{\bf{\it (ii)} $\mbox{\boldmath $\gamma\gamma$}$ production
 of $\mbox{\boldmath $(q\,\bar q)$}$
 pseudotensor-meson pairs}

\vspace{6pt}

Keeping only the physically relevant pieces (i.e. the terms
in $\beta^{2}\beta'^{2}$ of the series expansion in powers
of $\beta$, $\beta'$) of the helicity amplitudes
${\cal T}^{\Lambda\Lambda'}_{\lambda_\gamma\lambda'_\gamma}$,
we obtain~:

\begin{eqnarray}
 \left( {\cal T}^{0\:\!0}_{\pm\:\!\pm}\right)_{phys.} & = &
 \frac{K}{32}\,\frac{\langle\,(e_1-e_2)^2\rangle}{1-v^2}\,
 \frac{8a^2-3a+4b^2}{(a^2-b^2)^3}\,u^2u^{\prime 2}\beta^2
 \beta^{\prime 2} \; ,
 \label{pspp}
\end{eqnarray}

\begin{eqnarray}
 \left( {\cal T}^{0\:\!0}_{\pm\:\!\mp}\right)_{phys.} & = &
 \frac{K}{32}\,\frac{1}{(a^2-b^2)^3}\,\biggl\{\,
 \frac{\langle\,(e_1-e_2)^2\rangle}{1-v^2}\,
 (8a^2+4b^2-21a+9) -
 3\,\frac{\langle e_1e_2\rangle}{(a^2-v^2b^2)^5}
 \nonumber \\ &\times&\Bigl[\,
 a^5(4a^6-3a^5-16a^5b^2+15a^4b^2+20a^3b^4-20a^2b^4
 -8ab^6 \nonumber \\ &&\quad +\:8b^6)
 +4v^2a^3(-2a^8+3a^7+8a^7b^2-5a^6b^2-15a^5b^2-20a^5b^4
 \nonumber \\ &&\quad +\:25a^4b^4+25a^3b^4+26a^3b^6-45a^2b^6
 -10ab^6-10ab^8+20b^8) \nonumber \\ &&\quad
 +\: 2v^4a(-4a^9+70a^7b^2-70a^6b^4-165a^5b^4+8a^5b^6
 +181a^4b^6 \nonumber \\ &&\quad +\: 130a^3b^6-40a^3b^8
 -110a^2b^8-40ab^8+20ab^{10}+20b^{10}) \nonumber \\ &&\quad
 +\:4v^6b^2(-20a^8+30a^7b^2+65a^6b^2-8a^6b^4-79a^5b^4
 -75a^4b^4 \nonumber \\ &&\quad +\:20a^4b^6+65a^3b^6+35a^2b^6
 -10a^2b^8-23ab^8-2b^8+2b^{10}) \nonumber \\ &&\quad
 +\:v^8b^4(-40a^6+16a^5b^2+100a^4b^2-40a^3b^4-75a^2b^4
 +31ab^6 \nonumber \\ &&\quad +\:12b^6-4b^8)\,\Bigr]\biggr\}
 \,u^2u^{\prime 2}\beta^2\beta^{\prime 2}\; .
 \label{pspm}
\end{eqnarray}

Therefrom we get through Eq.~(\ref{mps}), with the choice
$\Phi_N(x)=\delta(x-1/2)$, $\Phi'_N(x')=\delta(x'-1/2)$~:

\begin{equation}
 {\cal M}^{0\:\!0}_{\pm\:\!\pm} = \frac{4}{225}\,K f_{2} f'_{2}
 \,\frac{\langle\,(e_{1}-e_{2})^2\rangle}{1-v^{2}} \; ,
 \label{mpspp}
\end{equation}

\begin{equation}
 {\cal M}^{0\:\!0}_{\pm\:\!\,\!\mp} = \frac{4}{225}\, K f_{2} f'_{2}
 \left\{\,\frac{\langle\,(e_{1}-e_{2})^2\rangle}{1-v^{2}} +
 6\,\langle e_{1}e_{2}\rangle\left[1-8v^{2}(1-v^{2})
 \right]\right\} \; .
 \label{mpspm}
\end{equation}

Finally, with the choice of the nonrelativistic
DA for the pseudotensor mesons,
Eq.~(\ref{dsdt}) leads us to~:

\begin{eqnarray}
 \frac{d\sigma^{\gamma\gamma\to QQ'}}{d(\cos\Theta)} & = &
 \frac{2^{19}\pi^{3}\alpha^{2}\alpha_{s}^{2} |f_{2}|^{2}
 |f_{2}'|^{2}\,\xi}{3^{6}5^{4}s^{3}}\,\biggl\{\,
 \frac{\langle\,(e_1-e_2)^2\rangle^{2}}{(1-v^2)^{2}}
 + 6\,\langle e_1e_2\rangle\langle\,(e_1-e_2)^2\rangle
 \,\frac{1-8v^{2}(1-v^{2})}{1-v^{2}} \nonumber \\ & + &
 18\,\langle e_1e_2\rangle^{2}\left[1-8v^{2}(1-v^{2})\right]^{2}
 \,\biggr\} \; .
 \label{dsgqps}
\end{eqnarray}

\noindent
{\bf{\it (iii)} $\mbox{\boldmath $\gamma\gamma$}$ production of
 a pion plus a $\mbox{\boldmath $(q\,\bar q)$}$ pseudotensor meson}

\vspace{6pt}

Let us identify $Q$ with the pion, $Q'$ with the pseudotensor
meson. Since for that process the expressions of the helicity
amplitudes
${\cal T}^{\Lambda\Lambda'}_{\lambda_\gamma\lambda'_\gamma}$
(even keeping only the physically relevant pieces, i.e.
the terms in $\beta^0\beta^{\prime 2}$ of the series expansion
in powers of $\beta$, $\beta'$) become particularly lengthy,
we shall not show them here. We shall thus restrict ourselves
to giving the expressions of the amplitudes
${\cal M}^{\Lambda\Lambda'}_{\lambda_\gamma\lambda'_\gamma}$
obtained with the choice
of the nonrelativistic DA for the pseudotensor meson
together with the Chernyak-Zhitnitsky DA for the pion,
as well as of the differential cross section derived
therefrom through Eq.~(\ref{dsdt})~:

\begin{equation}
 {\cal M}^{0\:\!0}_{\pm\:\!\pm} = \frac{1}{3\sqrt{3}}
 \,K f_{\pi} f'_{2}\,
 \frac{\langle\,(e_{1}-e_{2})^2\rangle}{1-v^{2}} \; ,
 \label{mpipp}
\end{equation}

\begin{eqnarray}
 {\cal M}^{0\:\!0}_{\pm\:\!\mp} & = & \frac{1}{3\sqrt{3}}\,
 K f_{\pi} f'_{2}\, \biggl\{ \frac{\langle\,(e_{1}-e_{2})^2\rangle}
 {1-v^{2}} + 3\,\frac{\langle e_{1}e_{2}\rangle}{2v^7}\,
 \Bigl[\,60v-106v^3+42v^5\nonumber\\&&\;
 +\,(30-63v^{2}+36v^{4}-4v^{6})
 \ln\left(\frac{1-v}{1+v}\right)\,\Bigr]\biggr\} \; ,
 \label{mpipm}
\end{eqnarray}

%Finally, with the above-defined choice of the nonrelativistic
%distribution amplitudes,
%Eq.~(\ref{dsdt}) leads us to~:

\begin{eqnarray}
 \frac{d\sigma^{\gamma\gamma\to QQ'}}{d(\cos\Theta)} & = &
 \frac{2^{14}\pi^{3}\alpha^{2}\alpha_{s}^{2} |f_{\pi}|^{2}
 |f_{2}'|^{2} \xi}{3^{5}s^{3}}\,\biggl\{\,
 2\,\frac{\langle\,(e_1-e_2)^2\rangle^{2}}{(1-v^2)^{2}}
 +6\,\frac{\langle e_1e_2\rangle\langle\,(e_1-e_2)^2\rangle}
 {2v^7}\nonumber\\&\times&\Bigl[60v-106v^{3}+42v^{5}+
 (30-63v^{2}+36v^{4}-4v^{6})\ln\left(\frac{1-v}{1+v}
 \right)\,\Bigr] \nonumber \\&+&
 9\,\frac{\langle e_1e_2\rangle^{2}}{4v^{14}}\,
 \Bigl[\,60v-106v^{3}+42v^{5}+(30-63v^{2}+36v^{4}-4v^{6})
 \ln\left(\frac{1-v}{1+v}\right)\,\Bigr]^{2}\,
 \biggr\} \; . \nonumber \\
 \label{dsgqpi}
\end{eqnarray}

\goodbreak

\setcounter{equation}{0}
\renewcommand{\theequation}{\mbox{C}\arabic{equation}}
\section*{ Appendix C~: Determination of the normalization
 constants $\mbox{\boldmath $|f_{L}|$}$}

\nobreak

The normalization constants $|f_{L}|$ for the various mesons $Q$
considered in this paper (except for the pion) are determined
through the experimental values of the corresponding decay widths
$\Gamma (Q \rightarrow \gamma \gamma)$,
using for the theoretical computation the formalism developed
in Sec. 2, but this time staying in the meson rest frame.

Let us define the helicity amplitude
${\cal M}^{LSJ\Lambda}_{\lambda_{\gamma}\lambda'_{\gamma}}(\Theta,\,\Phi)$
for the decay of a meson $Q$ with quantum numbers
$L,\,S,\,J,\,\Lambda$ into a pair of photons~;
here $\Theta$, $\Phi$ specify the direction
of one of the outgoing photons in the meson rest frame.
The dependence of
${\cal M}^{LSJ\Lambda}_{\lambda_{\gamma}\lambda'_{\gamma}}(\Theta,\,\Phi)$
on $\Theta$, $\Phi$ is known a priori \cite{bour}~:

\begin{equation}
 {\cal M}^{LSJ\Lambda}_{\lambda_{\gamma}\lambda'_{\gamma}}
 (\Theta,\,\Phi) = \hat{{\cal M}}_{\lambda_{\gamma}\lambda'_{\gamma}}
 \,d^J_{\Lambda,\lambda_{\gamma}-\lambda'_{\gamma}}(\Theta)
 \exp\left[i\Lambda\Phi\right]\; ,
 \label{mred}
\end{equation}

\noindent 
where the ``reduced'' amplitude
$\hat{{\cal M}}_{\lambda_{\gamma}\lambda'_{\gamma}}$
is independent of $\Lambda$, $\Theta$, and $\Phi$.
The decay width $\Gamma(Q\to\gamma\gamma)$ is then given as a function
of the reduced amplitudes
$\hat{{\cal M}}_{\lambda_{\gamma}\lambda'_{\gamma}}$
by the relation

\begin{equation}
 \Gamma(Q\to\gamma\gamma) = \frac{1}{160\pi M}
 \sum_{\lambda_{\gamma},\lambda'_{\gamma}}
 |\hat{{\cal M}}_{\lambda_{\gamma}\lambda'_{\gamma}}|^2 \; .
 \label{ggg}
\end{equation}

The equation to be used, in order to derive the helicity
amplitudes at the hadron level,
${\cal M}^{LSJ\Lambda}_{\lambda_{\gamma}\lambda'_{\gamma}}(\Theta,\,\Phi)$,
{}from the amplitudes
${\cal T}^{S\Lambda_{S}}_{\lambda_{\gamma}\lambda'_{\gamma}}
(\Theta,\,\Phi,\,\beta,\,\theta,\,\phi,\,x)$
defined for the corresponding parton process
($q\bar{q}\to\gamma\gamma'$), with $\beta$, $x$ defined as in
Sect. 2 and $\theta$, $\phi$ specifying the direction of the
incoming quark in the meson rest frame,
is quite similar to Eq.~(\ref{mfin})~:

\begin{eqnarray}
 {\cal M}^{LSJ\Lambda}_{\lambda_{\gamma}\lambda_{\gamma'}}
 (\Theta,\,\Phi) & = & f_{L} \lim_{\beta\to 0}
 \frac{1}{\beta^{L}}\int\frac{d(\cos\theta)}{2}
 \int\frac{d\phi}{2\pi}\sum_{\Lambda_{S}}
 \left[\zeta^{LSJ\Lambda_{L}\Lambda_{S}}(\theta,\,\phi)
 \right]^{\textstyle *}
 \nonumber \\ &\times& \int\frac{\Phi_{N}(x)\,dx}
 {\sqrt{x(1-x)}}\,{\cal T}^{S\Lambda_{S}}_{\lambda_{\gamma}
 \lambda'_{\gamma}}(\Theta,\,\Phi,\,\beta,\,\theta,\,\phi,\,x)\; ,
 \label{mgg}
\end{eqnarray}

\noindent
where $f_{L}$ and
$\zeta^{LSJ\Lambda_{L}\Lambda_{S}}(\theta,\,\phi)$
are given, respectively, in Eqs.~(\ref{fl}) [combined
with Eq.~(\ref{cl1})] and (\ref{zeta}).

We may choose the meson decay axis as our $z$-axis, thus
setting $\Theta = \Phi = 0$ and consequently
identifying the hadronic helicity amplitudes with
the corresponding ``reduced'' ones [see Eq.~(\ref{mred})]~;
on the other hand, since the reaction at parton level, as
seen in three-momentum space, occurs in a plane,
we may set $\phi=0$, so that Eq.~(\ref{mgg}) simplifies as
follows~:

\begin{eqnarray}
 \hat{{\cal M}}_{\lambda_{\gamma}\lambda_{\gamma'}}
 & = & f_{L} \lim_{\beta\to 0}
 \frac{1}{\beta^{L}}\int\frac{d(\cos\theta)}{2}\sum_{\Lambda_{S}}
 \zeta^{\Lambda_{L}\Lambda_{S}}(\theta,\,0)
 \nonumber \\ &\times& \int\frac{\Phi_{N}(x)\,dx}
 {\sqrt{x(1-x)}}\,{\cal T}^{\Lambda_{S}}_{\lambda_{\gamma}
 \lambda'_{\gamma}}(\beta,\,\theta,\,x)\; ,
 \label{msimpl}
\end{eqnarray}

\noindent
where we have dropped the superscripts $L$, $S$, $J$ from
$\zeta$ and $S$ from ${\cal T}$, and where we notice that
$\Lambda_{L}+\Lambda_{S}=\lambda_{\gamma}-\lambda'_{\gamma}$,
while $\theta$ is (except for an irrelevant change of sign)
the angle between incoming quark and outgoing photon
in the c.m. frame of the parton process.

In order to evaluate the helicity amplitudes at
parton level,
${\cal T}^{\Lambda_{S}}_{\lambda_{\gamma}\lambda'_{\gamma}}$,
we must consider the two Feynman graphs shown in Fig.~4.
Before  presenting the results in the particular cases of
interest, let us comment briefly on two problems
connected with the use of a running mass (depending on $x$)
for the quark and the antiquark.

\begin{enumerate}
\item[{\it i})] Gauge invariance is no longer preserved~;
however, it can be restored by adding a non-physical term
to the photon-quark coupling, i.e. by replacing, at each
photon-quark vertex, $\gamma^{\mu}$ by
$\gamma^{\mu}-[\slash{\kappa}\kappa^{\mu}/\kappa^2]_{\kappa^2\to 0}$,
where $\kappa$ is the four-momentum of the photon involved.
This way, each vertex becomes gauge-invariant by itself.
\item[{\it ii})] There is an ambiguity regarding the value
of the quark mass in the propagator of either diagram
of Fig.~4, since for that value one may choose  either
$m=M\sqrt{x^2-\beta^2/4}$ or $\bar{m}=M\sqrt{(1-x)^2-\beta^2/4}$
[see Eq.~(\ref{q0})].
However, as we have checked explicitly, both
choices give the same results for the helicity amplitudes
at the hadron level (i.e. after integration over $x$).
This is clearly related to the symmetry of the Feynman-graph
calculation and of $\Phi_N$ with respect to the exchange
$x \longleftrightarrow 1-x$.
\end{enumerate}
 
Notice that, when the nonrelativistic DA is used, both problems
are suppressed anyway.

Let us now consider separately the case of tensor and
pseudotensor mesons.

\vspace{6pt}

\noindent
{\bf {\it (i)} Tensor mesons}

\vspace{6pt}

In this case the helicity amplitudes at parton level,
${\cal T}^{\Lambda_{S}}_{\lambda_{\gamma}\lambda'_{\gamma}}$,
are given, up to the first order in the $\beta$ power expansion,
by the following expressions (amplitudes not given
explicitly here are vanishing)~:

\begin{equation}
 {\cal T}^{\,\pm 1}_{+\,+} = \pm\,\frac{1}{4}\,K'\,
 \frac{4x^2-6x+1}{x\sqrt{x(1-x)}}\,\beta\sin\theta\; ,
 \label{gt1pp}
\end{equation}

\begin{equation}
 {\cal T}^{\,\pm 1}_{-\,-} = {\cal T}^{\,\pm 1}_{+\,+}\; ,
 \label{gt1mm}
\end{equation}

\begin{equation}
 {\cal T}^{\,1}_{+\,-} =  -{\cal T}^{\,-1}_{-\,+} = 2 K'\,
 \frac{1-x}{\sqrt{x(1-x)}}\,\beta\sin\theta\; ,
 \label{gt1pm}
\end{equation}

\begin{equation}
 {\cal T}^{\,0}_{\pm\,\pm} = \frac{1}{2\sqrt{2}}\,K'\,
 \frac{1}{x\sqrt{x(1-x)}}\,\beta\cos\theta \; ,
 \label{gt0}
\end{equation}

\noindent
where $K'\!=\!8\sqrt{3}\,\pi\alpha\langle\,e_{q}^{2}\,\rangle$,
with $\langle\,e_{q}^{2}\,\rangle=5/(9\sqrt{2})$,
$1/(3\sqrt{2})$, and 1/9 respectively for $f_{2}(1270)$,
$a_{2}(1320)$, and $f'_{2}(1525)$.

Inserting Eqs.~(\ref{gt1pp})--(\ref{gt0}) into Eq.~(\ref{msimpl}),
taken for $L=S=1$, $J=2$,
we get the following expressions for the reduced hadronic
amplitudes $\hat{{\cal M}}_{\lambda_{\gamma}\lambda'_{\gamma}}$~:

\begin{equation}
 \hat{{\cal M}}_{\pm\,\pm} = \frac{1}{3\sqrt{3}}\,K'
 f_1 \int_0^1 dx \,\Phi_{N}(x)\,\frac{1-2x}{x^2} \; ,
 \label{gtpp}
\end{equation}

\begin{equation}
 \hat{{\cal M}}_{\pm\,\mp} = -\,\frac{2\sqrt{2}}{3}\,K'
 f_1 \int_0^1 dx\,\frac{\Phi_{N}(x)}{1-x}\; .
 \label{gtpm}
\end{equation}

Finally, using Eq.~(\ref{ggg}), and adopting a normalized
meson distribution amplitude $\Phi_{N}(x)$ of the
kind

\begin{equation}
 \Phi^{a}_{N}(x) = \frac{2^{2a+1}\Gamma(a+3/2)}{\sqrt{\pi}
 \Gamma(a+1)}\,x^a(1-x)^a \; ,
 \label{phias}
\end{equation}

\noindent
where $a \geq 1+L = 2$, we find

\begin{equation}
 \Gamma^{a}(Q\to\gamma\gamma) = \frac{16}{45}\,
 \frac{(2a+1)^2(6a^2-12a+7)}{a^2(a-1)^2}\,
 \pi\alpha^2\langle\,e_{q}^2\,\rangle^2\,\frac{1}{M}\,|f_1|^2 \; .
 \label{gta}
\end{equation}

In particular, for $a=2$, which means [see Eq.~(\ref{da23})]
using the generalized asymptotic DA for tensor mesons,
Eq.~(\ref{gta}) gives

\begin{equation}
 \Gamma(Q\to\gamma\gamma) = \frac{140}{9}\,\pi\alpha^2
 \langle\,e_{q}^2\,\rangle^2\,\frac{1}{M}\,|f_1|^2 \; ,
 \label{gta2}
\end{equation}

\noindent
while for $a\to \infty$ we recover the nonrelativistic DA
[i.e. $\Phi_{N}(x) = \delta(x-1/2)$], which gives in turn

\begin{equation}
 \Gamma(Q\to\gamma\gamma) = \frac{128}{15}\,\pi\alpha^2
  \langle\,e_{q}^2\,\rangle^2\,\frac{1}{M}\,|f_1|^2 \; .
 \label{gtanr}
\end{equation}

Making use of the available experimental data for the
masses and the two-photon decay widths of the
$f_2$, $a_2$, and $f'_2$ tensor mesons \cite{pdg} and of
Eqs.~(\ref{gta2}), (\ref{gtanr}), one gets estimates
of the absolute values of the corresponding $f_1$ constants.
The results of interest for the calculations of this paper are
presented in Tab. C1.

\vspace{6pt}

\goodbreak

\noindent
{\bf {\it (ii)} Pseudotensor mesons}

\vspace{6pt}

\nobreak

Here the only non-zero ones among the relevant helicity amplitudes
${\cal T}^{\Lambda_{S}}_{\lambda_{\gamma}\lambda'_{\gamma}}$
are given, up to order $\beta^2$,
by the following expression~:

\begin{equation}
 {\cal T}^{\,0}_{\pm\,\pm} = \pm\,\sqrt{2}\,K'\,
 \frac{1-x}{\sqrt{x(1-x)}}\,\mp\,\frac{1}{16\sqrt{2}}\,
 K'\,\frac{1-4(1-x)\cos^{2}\theta}{x^2(1-x)\sqrt{x(1-x)}}
 \,\beta^{2} \; .
 \label{gtps}
\end{equation}

Inserting Eq.~(\ref{gtps}) into Eq.~(\ref{msimpl}),
taken for $L=2$, $S=0$, $J=2$,
we get the following expression for the only non-vanishing
ones among the helicity amplitudes at the hadron level~:

\begin{equation}
 \hat{{\cal M}}_{\pm\,\pm} =
 \pm\,\frac{1}{30\sqrt{2}}\,K' f_{2} \int_0^1 dx \,
 \frac{\Phi_{N}(x)}{x^3(1-x)}\; .
 \label{gmps}
\end{equation}

Using again the distribution amplitude given in Eq.~(\ref{phias}),
but this time with $a\geq 1+L = 3$, and Eq.~(\ref{ggg}) we
find

\begin{equation}
 \Gamma^{a}(Q\to\gamma\gamma) =
 \frac{8}{375}\,\frac{(4a^2-1)^2}{a^2(a-2)^2}\,\pi\alpha^2
 \langle\,e_{q}^2\,\rangle^2\,\frac{1}{M}\,|f_2|^2\; .
 \label{gpsa}
\end{equation}

In particular, for $a=3$, which means [see Eq.~(\ref{da23})]
choosing the generalized asymptotic DA for pseudotensor
mesons, Eq.~(\ref{gpsa}) gives

\begin{equation}
 \Gamma(Q\to\gamma\gamma) =
 \frac{392}{135}\,\pi\alpha^2
 \langle\,e_{q}^2\,\rangle^2\,\frac{1}{M}\,|f_2|^2 \; ,
 \label{gpsa3} 
\end{equation}

\noindent
while for $a\to\infty$, which corresponds to the
nonrelativistic DA, we find

\begin{equation}
 \Gamma(Q\to\gamma\gamma) =
 \frac{128}{375}\,\pi\alpha^2
 \langle\,e_{q}^2\,\rangle^2\,\frac{1}{M}\,|f_2|^2 \; ,
 \label{gpsnr}
\end{equation}

\noindent
where, for $\pi_2(1670)$, $\langle\,e_{q}^2\,\rangle=1/(3\sqrt{2})$.

{}From these relations and from the available experimental data
\cite{pdg}
\footnote{\,It should be noticed that the value there given
for $\Gamma(\pi_2\to \gamma\gamma)$, namely 1.35 keV, which is
based on experimental data from the CELLO \cite{cello} and
Crystal Ball \cite{cball} Collaborations at DESY,
has very recently been contested by the ARGUS Collaboration
\cite{argus}~; this makes our numerical results
for pseudotensor-meson pair production, as well
as for hybrid pair production, somewhat more controversial.}
we can estimate the absolute value of the
$f_2$ constant for the $\pi_2$ pseudotensor meson.
The results are presented in Tab. C1.

\vspace{18pt}

\noindent{\Large\bf Acknowledgements}

\vspace{8pt}

Two of us (J.H. and F.M.) wish to thank the Laboratoire de Physique
Corpusculaire of the Coll\`ege de France for the warm hospitality
extended to them during a visit in summer 1996, when part of this
work was realized. This work has been partially supported by the
EU program ``Human Capital and Mobility'' under contract
CHRX-CT94-0450.

\newpage

\begin{center}
{\Large\bf Figure captions}
\end{center}
\vspace{10pt}
\begin{enumerate}
\item[{\bf Fig. 1}]
 Kinematic schemes for (A) the process $a b\to Q c$ in its
 center-of-mass frame~; (B) the process
 $a b \to q \bar{q} c$ in the c.m. frame of
 $q$ and $\bar{q}$ (meson rest frame).
\item[{\bf Fig. 2}]
 Differential cross section $E^8\,[d\sigma/dt]$ in 
 nb$\times$GeV$^{6}$, as a function of $\cos^2\Theta$,
 for the process $\gamma\gamma\to QQ'$ involving the
 production of tensor-meson pairs. The nonrelativistic
 DA [Eq.~(\ref{danr})] was used for the tensor mesons
 in Fig.~2(a), while the generalized asymptotic DA
 [Eq.~(\ref{da23})] was applied for them in Fig.~2(b).
 For comparison, analogous curves for $\gamma\gamma\to\pi\pi$,
 derived in the Brodsky-Lepage formalism, are also shown,
 using for the pion either the Chernyak-Zhitnitsky DA
 [Eq.~(\ref{dacz})] or the asymptotic DA [Eq.~(\ref{asypi})]
 (Figs.~2(a), 2(b) respectively).
\item[{\bf Fig. 3}]
 Differential cross section $E^8\,[d\sigma/dt]$ in
 nb$\times$GeV$^{6}$, as a function of $\cos^2\Theta$,
 for the process $\gamma\gamma\to QQ'$ involving the
 production of pseudotensor-meson and
 hybrid (one pion plus one pseudotensor meson) pairs.
 Both the asymptotic DA [Eq.~(\ref{asypi}), dashed curves]
 and the Chernyak-Zhitnitsky DA [Eq.~(\ref{dacz}), full curves]
 were used for the pion, while for the pseudotensor meson the
 nonrelativistic DA [Eq.~(\ref{danr})] was applied
 in Fig.~3(a) and the generalized asymptotic DA
 [Eq.~(\ref{da23})] in Fig.~3(b). 
\item[{\bf Fig. 4}]
 Lowest-order Feynman graphs for the process
 $q\bar q\to\gamma\gamma$.
\end{enumerate}

\newpage

\begin{center}
{\Large\bf Table captions}
\end{center}
\vspace{10pt}
\begin{enumerate}
\item[{\bf Tab. 1\phantom{A}}]
 Integrated cross sections (in $10^{-40}$ cm$^2$) of the
 process $ee'\to ee'QQ'$, involving the production of
 tensor-meson, pseudotensor-meson and hybrid (one pion
 plus one pseudotensor meson) pairs. For tensor and
 pseudotensor mesons, both the nonrelativistic DA
 [Eq.~(\ref{danr})] and the generalized asymptotic DA
 [Eq.~(\ref{da23})] were used (columns NR and GASY respectively);
 on the other hand both the asymptotic DA [Eq.~(\ref{asypi})] and
 the Chernyak-Zhitnitsky DA [Eq.~(\ref{dacz})] were considered
 for the pion. Experimental conditions here assumed are~:
 $\sqrt{s}=200$ GeV, $p_T > 1$ GeV.
\item[{\bf Tab. 2\phantom{A}}]
 Same as table 1, but assuming $p_T > 2$ GeV.
\item[{\bf Tab. 3\phantom{A}}]
 Same as table 1, but assuming~:
 $\sqrt{s}=10$ GeV, $p_T > 1$ GeV.
\item[{\bf Tab. B1}]
 Values of the charge factors $\langle\,(e_1-e_2)^2\rangle$
 and $\langle e_1e_2 \rangle$ for the meson pairs
 considered in this paper.
\item[{\bf Tab. C1}] Experimental two-photon decay widths,
 according to Ref.~\cite{pdg}, of the known $(q\,\bar{q})$
 tensor and pseudotensor mesons, and values of the normalization
 constant $|f_{L}|$ derived therefrom, using either the
 nonrelativistic DA [Eq.~(\ref{danr})] or the generalized
 asymptotic DA [Eq.~(\ref{da23})] (columns NR and GASY respectively).
\end{enumerate}

\newpage

\begin{figure}
\centerline{
\epsfig{figure=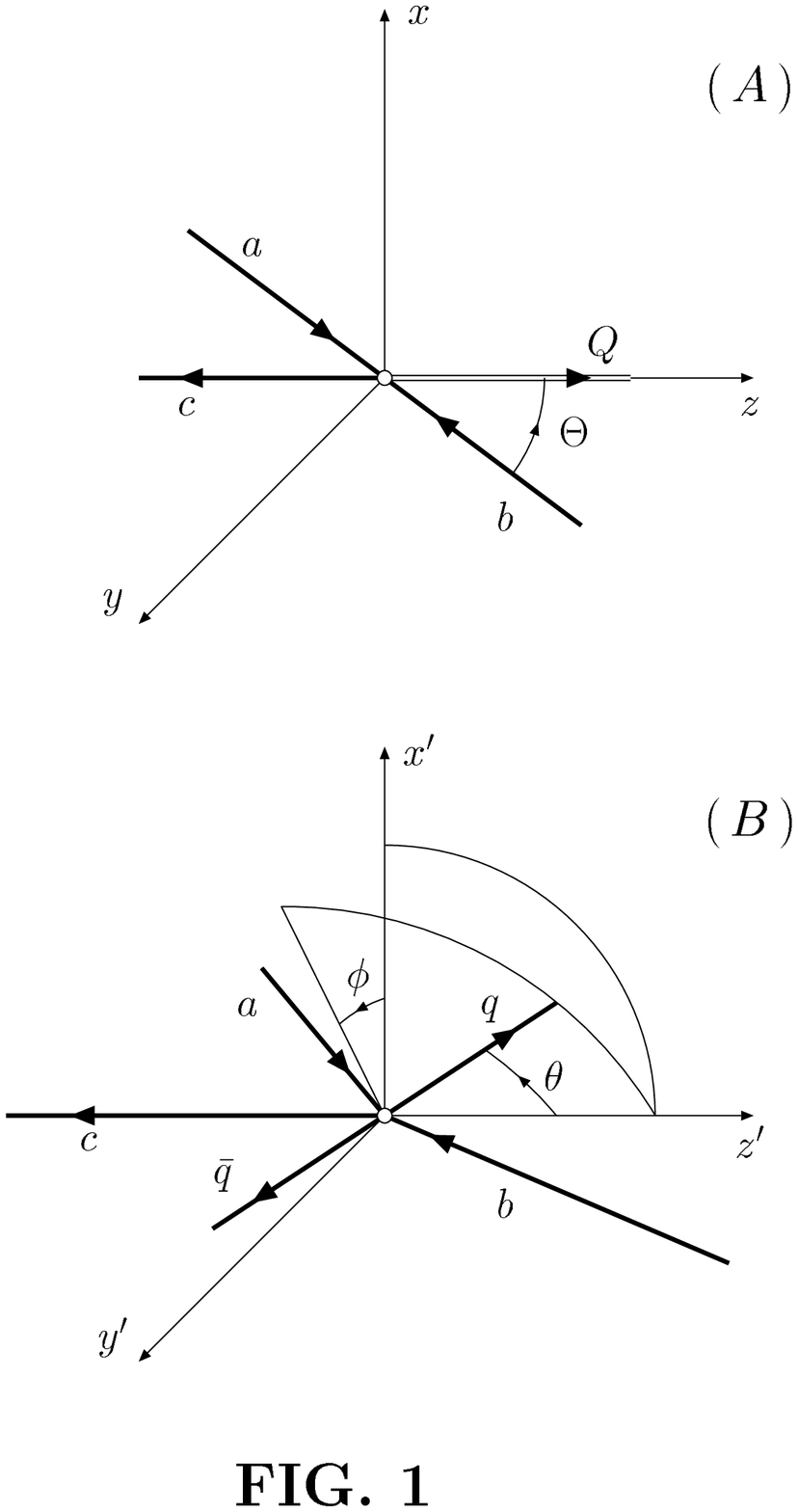,bbllx=50pt,bblly=120pt,bburx=500pt,%
bbury=780pt,width=14cm,height=20cm}}
\end{figure}

\clearpage

\begin{figure}
\centerline{
\epsfig{figure=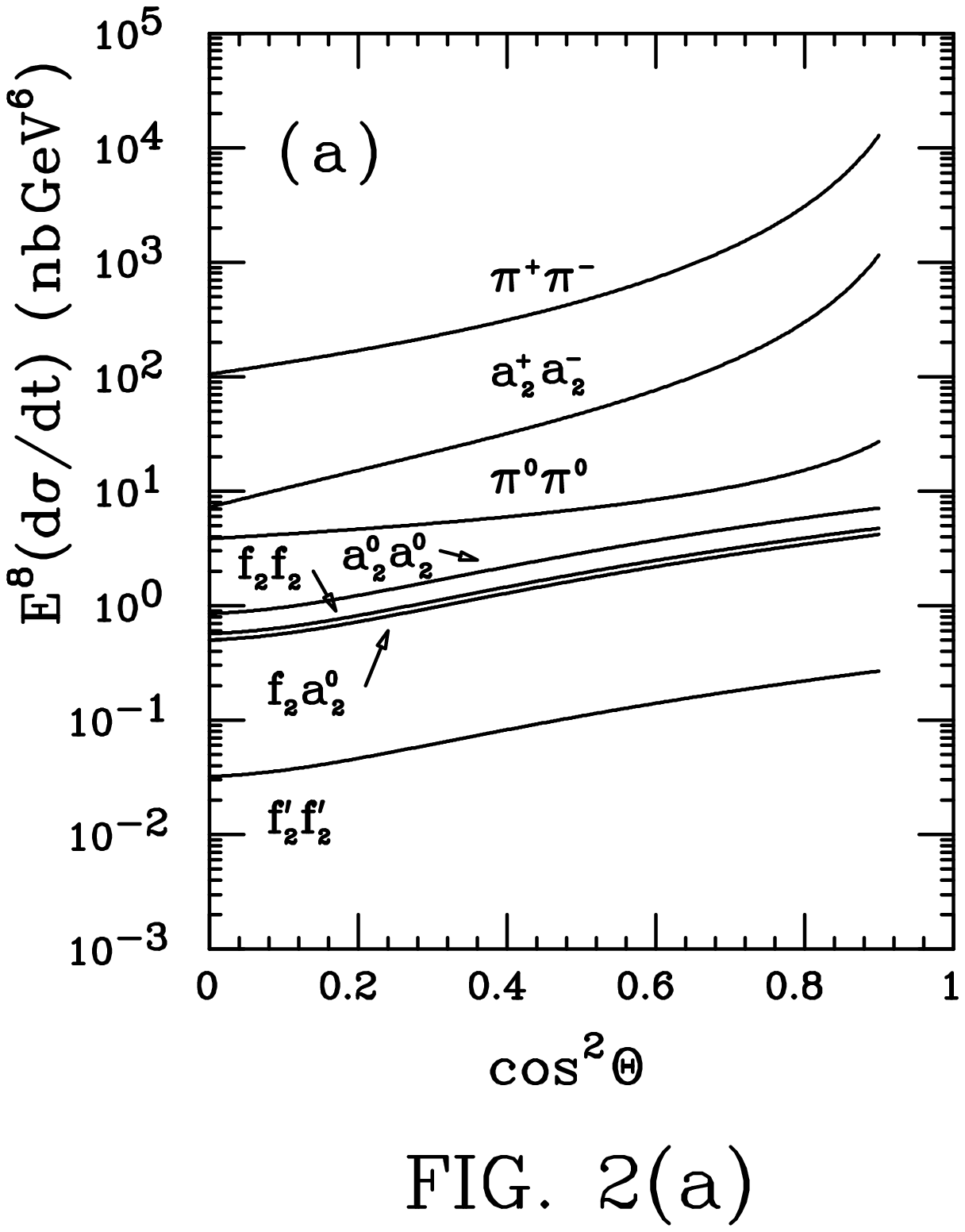,bbllx=90pt,bblly=100pt,bburx=460pt,%
bbury=610pt,width=14cm,height=18cm}}
\end{figure}

\clearpage

\begin{figure}
\centerline{
\epsfig{figure=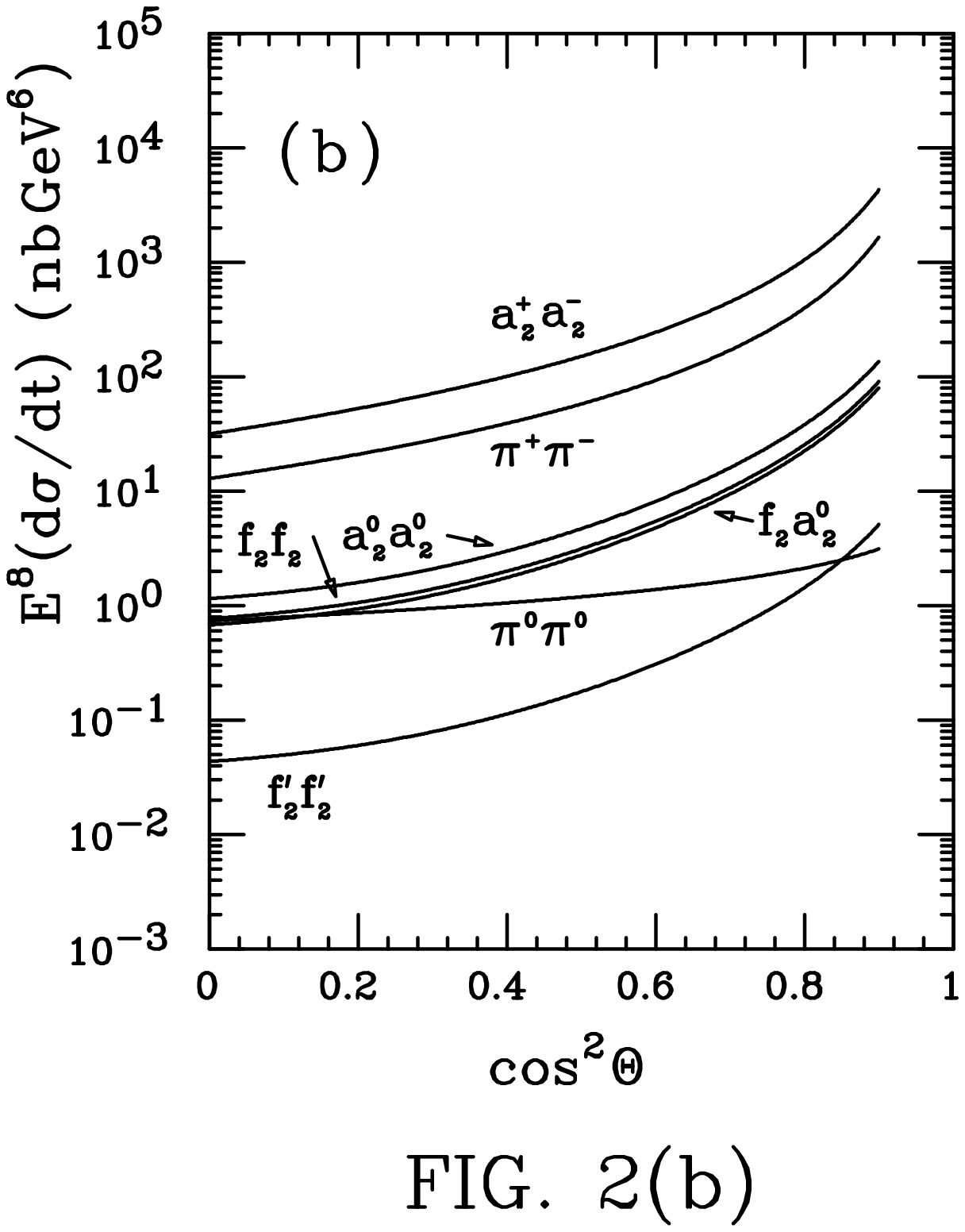,bbllx=90pt,bblly=100pt,bburx=460pt,%
bbury=610pt,width=14cm,height=18cm}}
\end{figure}

\clearpage

\begin{figure}
\centerline{
\epsfig{figure=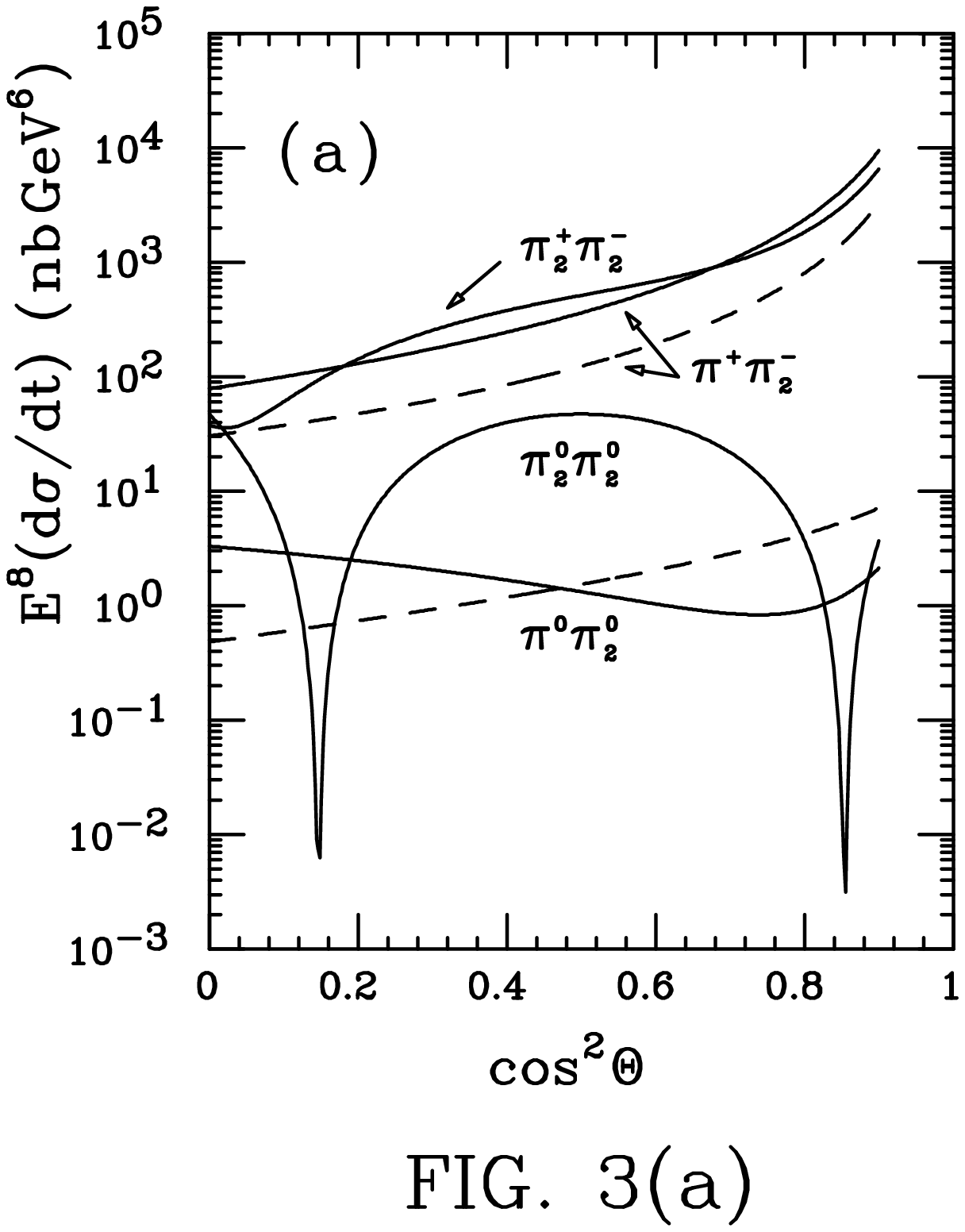,bbllx=90pt,bblly=100pt,bburx=460pt,%
bbury=610pt,width=14cm,height=18cm}}
\end{figure}

\clearpage

\begin{figure}
\centerline{
\epsfig{figure=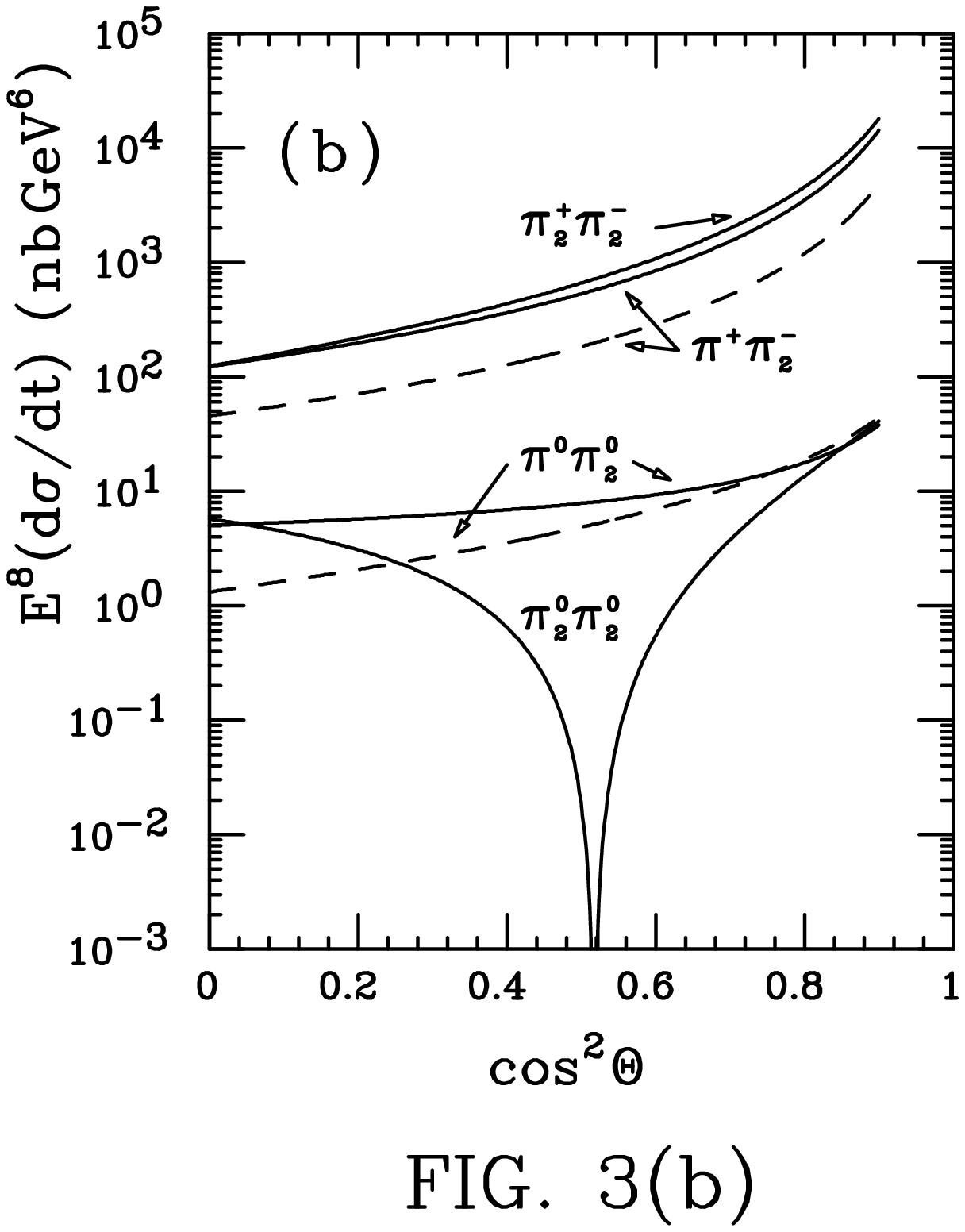,bbllx=90pt,bblly=100pt,bburx=460pt,%
bbury=610pt,width=14cm,height=18cm}}
\end{figure}

\clearpage

\begin{figure}
\centerline{
\epsfig{figure=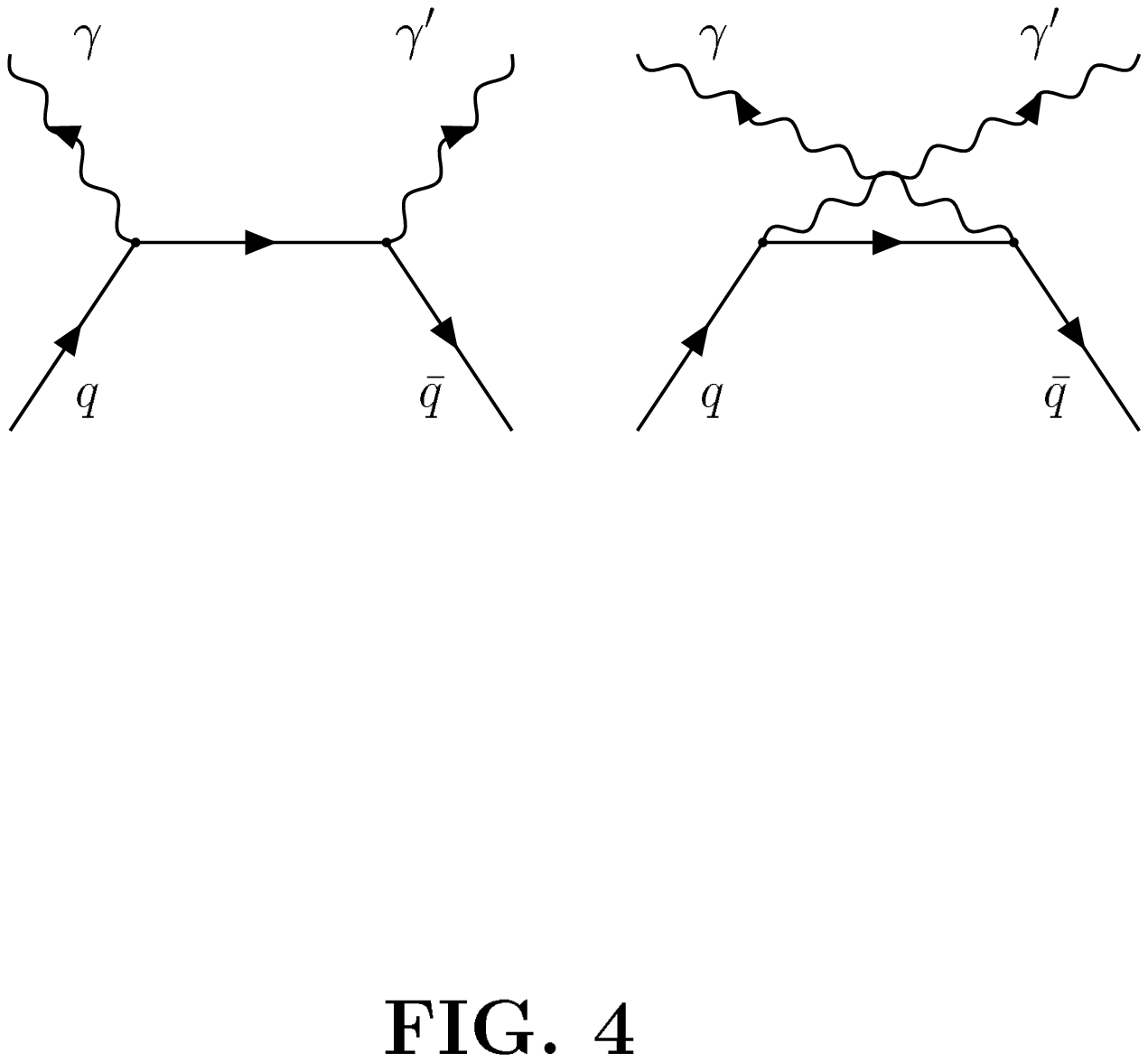,bbllx=50pt,bblly=160pt,bburx=500pt,%
bbury=600pt,width=14cm,height=18cm}}
\end{figure}

\clearpage

\begin{center}
 \begin{tabular}{crr}
  \hline\hline
  \noalign{\vspace{8pt}}
    $\quad\qquad QQ'\;\qquad$ &
    \multicolumn{2}{c}{$\quad\sigma(ee'\to ee'QQ')
                       \;[10^{-40}$ cm$^2]\quad$} \\
  \noalign{\vspace{4pt}}
  \cline{2-3}
  \noalign{\vspace{4pt}}
    & \multicolumn{1}{c}{$\quad$ NR} & \multicolumn{1}{c}{$\qquad$GASY}  \\
  \noalign{\vspace{4pt}}
  \hline
  \noalign{\vspace{8pt}}
    $f_2\,f_2$         &    20.9   &    25.4 $\qquad\quad$   \\
  \noalign{\vspace{4pt}} 
    $a^0_2\,a^0_2$     &    30.8   &    53.0 $\qquad\quad$   \\
  \noalign{\vspace{4pt}}
    $f_2\,a^0_2$       &    19.4   &    33.2 $\qquad\quad$   \\
  \noalign{\vspace{4pt}}
    $f'_2\,f'_2$       &     0.6   &     1.1 $\qquad\quad$   \\
  \noalign{\vspace{4pt}}
    $a^+_2\,a^-_2$     &   291.0   &  1003.3 $\qquad\quad$   \\
  \noalign{\vspace{4pt}}
    $\pi^0_2\,\pi^0_2$ &   141.1   &    23.2 $\qquad\quad$   \\
  \noalign{\vspace{4pt}}
    $\pi^+_2\,\pi^-_2$ &   749.3   &  1418.2 $\qquad\quad$   \\
  \noalign{\vspace{4pt}}
    $\pi^0_{\mbox{\tiny CZ}}\,\pi^0_2$   &   116.6   &
     263.0 $\qquad\quad$   \\
  \noalign{\vspace{4pt}}
    $\pi^0_{\mbox{\tiny ASY}}\,\pi^0_2$   &    35.2   &
     104.1 $\qquad\quad$   \\
  \noalign{\vspace{4pt}}
    $\pi^+_{\mbox{\tiny CZ}}\,\pi^-_2$   &  4167.0   &
     6562.6 $\qquad\quad$   \\
  \noalign{\vspace{4pt}}
    $\pi^+_{\mbox{\tiny ASY}}\,\pi^-_2$   &  1495.8   &
     2368.5 $\qquad\quad$   \\
  \noalign{\vspace{8pt}}
  \hline\hline
  \noalign{\vspace{16pt}}
  \multicolumn{3}{c}{\Large Table 1}
 \end{tabular}
\end{center}

\vspace{0.5truecm}

\begin{center}
 \begin{tabular}{crr}
  \hline\hline
  \noalign{\vspace{8pt}}
    $\quad\qquad QQ'\;\qquad$ &
    \multicolumn{2}{c}{$\quad\sigma(ee'\to ee'QQ')
                       \;[10^{-40}$ cm$^2]\quad$} \\
  \noalign{\vspace{4pt}}
  \cline{2-3}
  \noalign{\vspace{4pt}}
    & \multicolumn{1}{c}{$\quad$ NR} & \multicolumn{1}{c}{$\qquad$GASY}  \\
  \noalign{\vspace{4pt}}
  \hline
  \noalign{\vspace{8pt}}
    $f_2\,f_2$         &     1.0  &      0.9 $\qquad\quad$   \\
  \noalign{\vspace{4pt}} 
    $a^0_2\,a^0_2$     &     1.9   &     2.8 $\qquad\quad$   \\
  \noalign{\vspace{4pt}}
    $f_2\,a^0_2$       &     1.2   &     1.7 $\qquad\quad$   \\
  \noalign{\vspace{4pt}}
    $f'_2\,f'_2$       &    0.05   &    0.08 $\qquad\quad$   \\
  \noalign{\vspace{4pt}}
    $a^+_2\,a^-_2$     &    16.4   &    55.6 $\qquad\quad$   \\
  \noalign{\vspace{4pt}}
    $\pi^0_2\,\pi^0_2$ &    17.5   &     3.4 $\qquad\quad$   \\
  \noalign{\vspace{4pt}}
    $\pi^+_2\,\pi^-_2$ &    72.3   &   126.2 $\qquad\quad$   \\
  \noalign{\vspace{4pt}}
    $\pi^0_{\mbox{\tiny CZ}}\,\pi^0_2$    &     5.7   &
     11.5 $\qquad\quad$   \\
  \noalign{\vspace{4pt}}
    $\pi^0_{\mbox{\tiny ASY}}\,\pi^0_2$    &     1.4   &
     4.1 $\qquad\quad$   \\
  \noalign{\vspace{4pt}}
    $\pi^+_{\mbox{\tiny CZ}}\,\pi^-_2$    &   156.7   &
     246.5 $\qquad\quad$   \\
  \noalign{\vspace{4pt}}
    $\pi^+_{\mbox{\tiny ASY}}\,\pi^-_2$    &    56.0   &
     88.7 $\qquad\quad$   \\
  \noalign{\vspace{8pt}}
  \hline\hline
  \noalign{\vspace{16pt}}
  \multicolumn{3}{c}{\Large Table 2}
 \end{tabular}
\end{center}

\newpage

\begin{center}
 \begin{tabular}{crr}
  \hline\hline
  \noalign{\vspace{8pt}}
    $\quad\qquad QQ'\;\qquad$ &
    \multicolumn{2}{c}{$\quad\sigma(ee'\to ee'QQ')
                       \;[10^{-40}$ cm$^2]\quad$} \\
  \noalign{\vspace{4pt}}
  \cline{2-3}
  \noalign{\vspace{4pt}}
    & \multicolumn{1}{c}{$\quad$ NR} & \multicolumn{1}{c}{$\qquad$GASY}  \\
  \noalign{\vspace{4pt}}
  \hline
  \noalign{\vspace{8pt}}
    $f_2\,f_2$         &     1.4   &     1.4 $\qquad\quad$   \\
  \noalign{\vspace{4pt}} 
    $a^0_2\,a^0_2$     &     2.3   &     3.6 $\qquad\quad$   \\
  \noalign{\vspace{4pt}}
    $f_2\,a^0_2$       &     1.5   &     2.3 $\qquad\quad$   \\
  \noalign{\vspace{4pt}}
    $f'_2\,f'_2$       &    0.04   &    0.06 $\qquad\quad$   \\
  \noalign{\vspace{4pt}}
    $a^+_2\,a^-_2$     &    20.1   &    68.6 $\qquad\quad$   \\
  \noalign{\vspace{4pt}}
    $\pi^0_2\,\pi^0_2$ &    10.0   &     1.6 $\qquad\quad$   \\
  \noalign{\vspace{4pt}}
    $\pi^+_2\,\pi^-_2$ &    41.6   &    72.4 $\qquad\quad$   \\
  \noalign{\vspace{4pt}}
    $\pi^0_{\mbox{\tiny CZ}}\,\pi^0_2$    &    12.0   &
     24.5 $\qquad\quad$   \\
  \noalign{\vspace{4pt}}
    $\pi^0_{\mbox{\tiny ASY}}\,\pi^0_2$    &     3.0   &
     8.8 $\qquad\quad$   \\
  \noalign{\vspace{4pt}}
    $\pi^+_{\mbox{\tiny CZ}}\,\pi^-_2$    &   341.1   &
     536.7 $\qquad\quad$   \\
  \noalign{\vspace{4pt}}
    $\pi^+_{\mbox{\tiny ASY}}\,\pi^-_2$    &   122.0   &
     193.2 $\qquad\quad$   \\
  \noalign{\vspace{8pt}}
  \hline\hline
  \noalign{\vspace{20pt}}
  \multicolumn{3}{c}{\Large Table 3}
 \end{tabular}
\end{center}

\vspace{1.5truecm}

\begin{center}
 \begin{tabular}{ccc}
  \hline\hline
  \noalign{\vspace{8pt}}
    $\quad Q\,Q'\quad $ & $\quad\langle\,(e_1-e_2)^2\rangle\quad$ &
    $\quad\langle e_1e_2 \rangle\quad$ \\
  \noalign{\vspace{8pt}}
  \hline
  \noalign{\vspace{8pt}}
    $f_2\,f_2$         & 0 & 5/18   \\
  \noalign{\vspace{4pt}}
    $a^0_2\,a^0_2$     & 0 & 5/18   \\
  \noalign{\vspace{4pt}}
    $f_2\,a^0_2$       & 0 & 1/6    \\
  \noalign{\vspace{4pt}}
    $f'_2\,f'_2$       & 0 & 1/9    \\
  \noalign{\vspace{4pt}}
    $a^+_2\,a^-_2$     & 1 & $-2/9$ \\
  \noalign{\vspace{4pt}}
    $\pi^0_2\,\pi^0_2$ & 0 & 5/18   \\
  \noalign{\vspace{4pt}}
    $\pi^+_2\,\pi^-_2$ & 1 & $-2/9$ \\
  \noalign{\vspace{4pt}}
    $\pi^0\,\pi^0_2$   & 0 & 5/18   \\
  \noalign{\vspace{4pt}}
    $\pi^+\,\pi^-_2$   & 1 & $-2/9$ \\
  \noalign{\vspace{8pt}}
  \hline\hline
  \noalign{\vspace{20pt}}
  \multicolumn{3}{c}{\Large Table B1}
 \end{tabular}
\end{center}

\newpage

\vspace{36pt}

\begin{center}
 \begin{tabular}{ccrr}
  \hline\hline
  \noalign{\vspace{8pt}}
    $\quad Q\quad$ &
    \multicolumn{1}{c}{$\quad\Gamma(Q\to\gamma\gamma)$ (keV)$\quad$} &
    \multicolumn{2}{c}{$\quad|f_{L}|\,$ (MeV)$\quad$} \\
  \noalign{\vspace{4pt}}
  \cline{3-4}
  \noalign{\vspace{4pt}}
    & & \multicolumn{1}{c}{NR} & \multicolumn{1}{c}{$\quad$GASY}  \\
  \noalign{\vspace{4pt}}
  \hline
  \noalign{\vspace{8pt}}
    $f_2(1270)$   &   2.440   &   119   &   88$\quad$  \\
  \noalign{\vspace{4pt}}
    $a_2(1320)$   &   1.040   &   131   &   97$\quad$  \\
  \noalign{\vspace{4pt}}
    $f'_2(1525)$  &   0.097   &    92   &   68$\quad$  \\
  \noalign{\vspace{4pt}}
    $\pi_2(1670)$ &   1.350   &   845   &  290$\quad$  \\
  \noalign{\vspace{8pt}}
  \hline\hline
  \noalign{\vspace{20pt}}
  \multicolumn{4}{c}{\Large Table C1}
 \end{tabular}
\end{center}


\newpage

\begin{thebibliography}{99}

\bibitem{brle}	G.P.~Lepage, S.J.~Brodsky~: Phys. Rev. D22,
		2157 (1980)~; S.J.~Brodsky, G.P.~Lepage~:
		{\it ibid.} 24, 1808 (1981)~;
		24, 2948 (1981).
\bibitem{brod}	See, e.g., S.J.~Brodsky, in Proc. of the IX International
		Workshop on Photon-Photon Collisions, San Diego 1992,
		eds. D.O. Caldwell and H.P. Paar (World Scientific,
		Singapore, 1992), p. 209.
\bibitem{cahn}	R.N.~Cahn~: Phys. Rev. D35, 3342 (1987).
\bibitem{kada}	E.H.~Kada, P.~Kessler, J.~Parisi~:
		Phys. Rev. D39, 2657 (1989).
\bibitem{yaou}	L.~Houra-Yaou, P.~Kessler, J.~Parisi~:
		Phys. Rev. D45, 794 (1992).
\bibitem{icho}	A.~Ichola, J.~Parisi~:
		Z. Phys. C66, 653 (1995).
\bibitem{murg}	F.~Murgia, P.~Kessler, J.~Parisi~:
		Z. Phys. C71, 483 (1996).
\bibitem{bena}	M.~Benayoun, M.~Froissart~:
		Nucl. Phys. B315, 295 (1989).
\bibitem{cher}	V.L.~Chernyak, A.R.~Zhitnitsky~:
		Nucl. Phys. B201, 492 (1982).
\bibitem{mack}	S.J.~Brodsky, G.P.~Lepage, P.B.~Mackenzie~:
		Phys. Rev. D28, 228 (1983).
\bibitem{crji}	C.R.~Ji, F.~Amiri~:
		Phys. Rev. D42, 3764 (1990).
\bibitem{bour}	See, e.g., C.~Bourrely, E.~Leader, J.~Soffer~:
		Phys. Rep. 59, 95 (1980).
\bibitem{pdg}	Particle Data Group, R.M.~Barnett {\it et al.}~:
		Phys. Rev. D54, 1 (1996).
\bibitem{cello}	CELLO Collaboration, H.J.~Behrend {\it et al.}~:
		Z. Phys. C46, 583 (1990).
\bibitem{cball}	Crystal Ball Collaboration, D.~Antreasyan
		{\it et al.}~: Z. Phys. C48, 561 (1990).
\bibitem{argus}	ARGUS Collaboration, H.~Albrecht {\it et al.}~:
		Report DESY 96-112, June 1996.

\end{thebibliography}
\end{document}